\documentclass[a4paper,11pt]{article}

\usepackage{jheppub}
\usepackage[utf8]{inputenc}
\usepackage{slashed,cancel}
\usepackage[section]{placeins}
\usepackage{bm}
\usepackage{pifont}
\usepackage{tabularx}  
\usepackage{adjustbox} 
\usepackage[english]{babel}
\usepackage{multirow}
\usepackage[normalem]{ulem} 
\usepackage{booktabs}
\usepackage{braket}
\usepackage{MnSymbol}
\usepackage{graphicx}
\usepackage{float}
\usepackage{subcaption}
\usepackage{multirow}
\usepackage[skins,theorems]{tcolorbox}
\usepackage[capitalise]{cleveref}
\usepackage{orcidlink}

\tcbset{highlight math style={enhanced,
  colframe=red,colback=white,arc=0pt,boxrule=1pt}}

\renewcommand{\d}{\text{d}}

\numberwithin{equation}{section}

\definecolor{rosso}{cmyk}{0,1,1,0.4}
\definecolor{rossos}{cmyk}{0,1,1,0.55}
\definecolor{rossoc}{cmyk}{0,1,1,0.2}
\definecolor{blu}{cmyk}{1,1,0,0.3}
\definecolor{blus}{cmyk}{1,1,0,0.6}
\definecolor{bluc}{cmyk}{1,1,0,0.1}
\definecolor{verde}{cmyk}{0.92,0,0.59,0.25}
\definecolor{verdec}{cmyk}{0.92,0,0.59,0.15}
\definecolor{verdes}{cmyk}{0.92,0,0.59,0.4}

\newcommand\subsetsim{\mathrel{%
  \ooalign{\raise0.2ex\hbox{$\subset$}\cr\hidewidth\raise-0.8ex\hbox{\scalebox{0.9}{$\sim$}}\hidewidth\cr}}}
\newcommand*{\diff}[1]{\text{d}#1}

\graphicspath{{Draft_Plots/}}

\begin{document}

\title{\Large LHC Constraints on Resonant Kaluza-Klein Gravitons}

\author[]{Arturo de Giorgi \orcidlink{0000-0002-9260-5466
},}
\author[]{Matteo Marcoli \orcidlink{0000-0002-9405-1399},}
\author[]{Federico Silvetti \orcidlink{0000-0002-9135-2364}}

\affiliation[]{Institute for Particle Physics Phenomenology, Department of Physics,\\ Durham University, Durham DH1 3LE, U.K.}

\emailAdd{arturo.de-giorgi@durham.ac.uk}
\emailAdd{matteo.marcoli@durham.ac.uk}
\emailAdd{federico.silvetti@durham.ac.uk}

\preprint{IPPP/26/31}

\abstract{
The signature prediction of extra-dimensional theories is the appearance of a tower of massive gravitons. In this work, we study the constraints on resonant heavy spin-2 particles at the Large Hadron Collider (LHC), with focus on the entire tower of gravitons, beyond the traditional single-resonance analysis. Since several states of the tower can lie within the same accessible mass window, the combined signal is enhanced, and the resulting constraints can be significantly stronger than those obtained from a single resonance alone. We first update the current constraints on single graviton searches stemming from diphoton and dilepton data from ATLAS and CMS, using datasets collected above $\sim 200$~GeV, and then consider the impact on the bounds of the full tower of states for different extra-dimensional scenarios. This allows us to extend the reach of current searches to lower mass regions than typically considered, paving the way for future analyses by experimental collaborations.
}

\maketitle 
\flushbottom 

\section{Introduction}
Since the discovery of the Higgs boson~\cite{ATLAS:2012yve,CMS:2012qbp}, the Standard Model~(SM) has been regarded as the renormalisable, ultraviolet‑complete theory that successfully describes three of the four known fundamental interactions. Gravity remains an exception, being still described by the non‑renormalisable framework of General Relativity~(GR). Nonetheless, GR has, to date, passed all experimental tests with remarkable precision~\cite{Will:2014kxa}.
From the QFT point of view, GR is the leading-order effective field theory~(EFT) of a propagating massless spin-2 particle, the graviton.
The unique combination of quadratic operators that endows the graviton with a mass, and allows for the propagation of the correct five degrees of freedom without introducing ghosts, was identified long ago by Fierz and Pauli~\cite{Fierz:1939ix}. Current constraints from gravitational‑wave dispersion and from solar‑system planetary dynamics place upper bounds on the GR graviton mass at the level of $\sim 10^{-23}$~eV~\cite{LIGOScientific:2020tif,Bernus:2020szc}.

The question of whether a deeper theoretical reason compels the graviton to be massless, and whether additional massive spin‑2 fields could exist in Nature, has driven the quest for a consistent massive theory of gravity. This effort is highly non‑trivial, since generic non‑linear combinations of graviton self‑interactions typically give rise to instabilities. 
The unique ghost-free potential for a theory of massive gravity was identified only within the last $\sim15$~years, giving rise to what is known today as ``dRGT massive gravity''~\cite{deRham:2010ik,deRham:2010kj}. Shortly after, this construction was extended to a fully dynamical two-metric theory, commonly referred to as (ghost-free) bimetric gravity~\cite{Hassan:2011zd,Hassan:2011vm,deRham:2013tfa}.
Constraints on single massive gravitons were derived in a variety of observables in the literature, ranging from colliders to astrophysics and cosmology~\cite{Cembranos:2017vgi,Armaleo:2019gil,Armaleo:2020yml,Cembranos:2021vdv,dEnterria:2023npy,Voronchikhin:2023qig,Voronchikhin:2024ygo,Garcia-Cely:2025ula,Voronchikhin:2025eqm,Cembranos:2026cqk,Gue:2026wqd}.
The presence of a single massive graviton, however, introduces two further challenges. First, the longitudinal components of the massive graviton grow with energy and lower significantly the EFT cutoff compared to what is naively expected from the Lagrangian~\cite{Arkani-Hamed:2002bjr,Schwartz:2003vj,Bonifacio:2019mgk,Falkowski:2020mjq}. Second, a theory containing only one massive graviton violates positivity bounds unless the effective cutoff lies within a factor of $\mathcal{O}(10)$ of the graviton mass~\cite{Bellazzini:2023nqj,Dong:2025dpy}, thereby becoming a poor EFT and needing soon a UV completion.

The above limitations and issues can be lifted if the theory features multiple gravitons close enough in mass scale. While this may seem an ad-hoc construction from a bottom-up perspective, it is naturally realised in extra-dimensional theories of gravity. The most solid prediction of compact extra dimensions is the appearance of a tower of massive gravitons in the resultant four-dimensional EFT, usually called ``Kaluza-Klein"~(KK) tower.
Restoring unitarity at high energies involves the inclusion of the full tower in the computations. The exact mechanism of the restoration depends on the geometry of the extra dimensions~\cite{Schwartz:2003vj,Bonifacio:2019ioc}. Explicit model-dependent verifications have been presented in the context of flat and warped extra-dimensional models~\cite{Chivukula:2020hvi,deGiorgi:2020qlg,Chivukula:2023qrt,Chivukula:2023sua,deGiorgi:2023mdy}.

Since the seminal works of Kaluza and Klein~(KK)~\cite{Kaluza:1921tu,Klein:1926tv}, extra-dimensional models have found a revival in the early 2000s due to their ability to elegantly solve hierarchy puzzles~\cite{Antoniadis:1990ew,Arkani-Hamed:1998jmv,Antoniadis:1998ig,Randall:1999ee,Randall:1999vf}. 
Extra-dimensional models come with a rich set of phenomenology related to open experimental problems of the SM: they can act as natural gravitational portal to dark matter and dark sectors in general~\cite{deGiorgi:2021xvm,deGiorgi:2022yha,Voronchikhin:2023znz,Chen:2023oip,Voronchikhin:2023qig,Koutroulis:2024wjl,Chivukula:2024nzt,Lee:2024wes,Donini:2025cpl,Donini:2025qrf,Clarke:2026cdl}, they can participate in generating neutrino masses~\cite{Dienes:1998sb,Arkani-Hamed:1998wuz,Dvali:1999cn,Lukas:2000rg,Grossman:1999ra,Huber:2003sf,Fong:2011xh,deGiorgi:2025xgp}, or be used as frameworks for different types of new physics~(NP), axions being nowadays among the most popular candidates~\cite{Dienes:1999gw,DiLella:2000dn,Choi:2003wr,Flacke:2006ad,deGiorgi:2024elx,Craig:2024dnl,Reece:2025thc,Choi:2026kxu,Albertus:2026fbe,Arza:2026rsl}.

Such interest prompts an in-depth study of the available parameter space of such models across multiple energy scales. For massive gravitons heavier than $\sim 200$~GeV, the LHC is currently the only active source of data.
Over the past decades, there has been an intense search for resonant heavy gravitons both at Tevatron~\cite{D0:2005srl,CDF:2010muc} and LHC in the diphoton and dilepton channels~\cite{ATLAS:2011ab,ATLAS:2011tzr,CMS:2011bsw,ATLAS:2012hvw,ATLAS:2015shg,CMS:2016kgr,ATLAS:2019erb,CMS:2021ctt,ATLAS:2021uiz,CMS:2024nht}, as well as in massive gauge bosons and di-Higgs channels~\cite{ATLAS:2017uhp,CMS:2019qem,CMS:2019kaf,ATLAS:2020tlo,CMS:2021roc}. However, typical experimental analysis focuses on single-graviton searches. This is motivated by the fact that constraints get weaker for larger masses, and thus one expects the contribution from heavier KK modes to be subleading. This is only partially true. If the mass of the lightest graviton is sufficiently small, the number of relevant resonances increases. This can partially compensate for the decreasing relevance of heavy resonances. To the best of our knowledge, steps in this direction were taken solely in Refs.~\cite{ATLAS:2023hbp,Bhattacharya:2024wpy}, 
limited to the context of clockwork constructions for extra-dimensional gravitons~\cite{Giudice:2016yja} and axion-like particles~\cite{Choi:2014rja,Choi:2015fiu,Kaplan:2015fuy}. 

In this work, we would like to take a step further in pushing the traditional resonant analysis. In particular, we aim (i) at generalising it to a more comprehensive set of constraints, including different channels and extra-dimensional models, and (ii) at extending the constraints to graviton masses smaller than the Higgs scale, where typically these searches are not performed. We hope in this way to guide experimental collaborations toward systematically extending current analyses, while instructing phenomenologists on how single-graviton constraints should be treated when compared to the full picture.

\bigskip
The structure of the paper is the following. In Sec.~\ref{sec:models} we present the model and discuss the limitations stemming from its EFT nature. In Sec.~\ref{sec:method}, we present the data and the procedure that will be implemented for the derivation of the constraints. 
We then apply it in Sec.~\ref{sec:analysis} first to the single graviton case, where we discuss the current constraints stemming from diphoton and dilepton production measurements by the ATLAS and CMS experiments, and secondly to the extra-dimensional case, highlighting the modifications and the limitations of the analysis, and extending the constraints to masses lower than the Higgs scale. Finally, we draw our conclusions and comment on future prospects in Sec.~\ref{sec:conclusions}.
\section{Benchmark Models}
\label{sec:models}

The effective Lagrangian describing the interaction of a set of $N$ massive gravitons with the SM at leading order in the coupling expansion reads
\begin{equation}\label{eq:interactionLagrangian}
    \mathcal{L}_\text{int.}= -\left[\sum\limits_{n=1}^N \frac{1}{{\Lambda_n}}G^{(n)}_{\mu\nu}\right]T^{\mu\nu}\equiv -\frac{1}{{\Lambda}}\left[\sum\limits_{n=1}^N c_nG^{(n)}_{\mu\nu}\right]T^{\mu\nu}\,,
\end{equation}
where $T^{\mu\nu}$ is the energy-momentum tensor of the SM, $G^{(n)}_{\mu\nu}$ is a massive graviton field, $\Lambda$ is the overall interaction scale, and  $c_n$ are model-dependent dimensionless coefficients. We will denote by $m_n$ the masses of the corresponding gravitons, and for the single-graviton case, we will identify $N=1$, $c_1=1$, and $m_G\equiv m_1$. 

The interaction scale $\Lambda$ dictates the strength of the couplings of gravitons to matter, but the effective cutoff scale of the theory $\Lambda_\text{eff.}$ can be sensibly smaller than $\Lambda$. Indeed, the scattering amplitudes of the longitudinal graviton modes grow with additional powers of the energy, thereby lowering the effective cutoff scale of the theory. This is in analogy with the growth of longitudinal massive gauge–boson scattering in the SM when the Higgs boson is removed. Denoting
\begin{equation}
    \Lambda_\alpha \equiv \left(m_1^{\alpha-1} \Lambda\right)^{\frac{1}{\alpha}}\,,
\end{equation}
the cutoff for a single massive graviton is of order $\Lambda_3$, whereas in the single extra‑dimensional setup it is raised to $\Lambda_{3/2}$~\cite{Schwartz:2003vj}. Throughout our analysis, we require the cutoff scale to exceed the mass of the heaviest graviton included in the analysis. For a single massive graviton, this simply implies $\Lambda > m_G$. 
In extra‑dimensional constructions, however, the cutoff condition becomes non‑trivial and depends on the ratio between the heaviest and lightest KK modes.

\bigskip
We consider as motivated working phenomenological benchmark the case in which all massive gravitons interact equally, and their masses are approximately equally spaced
\begin{align}
    &m_n = n\times m_1\,, &c_n=1\,.
\end{align}
This is the simplest benchmark to test the effects of a tower of states, as their density is constant. Nevertheless, it can also be found in realistic models.
Indeed, it matches the 5-dimensional Randall-Sundrum~(RS) model~\cite{Randall:1999ee,Randall:1999vf} in the low warping limit, and provides a good approximation at large warping. For completeness, we also report the constraints on the exact RS spectrum in App.~\ref{app:RS-model}.

The tower of gravitons needs to be cut off at some value $n=N>1$ in order to preserve the validity of the EFT.
Since we will consider bins up to $6$~TeV, overall EFT consistency demands an effective cutoff larger than the heaviest graviton, $m_N$, to $\Lambda_{\text{eff}} > m_N> 6$~TeV.
The cutoff condition then translates to
\begin{equation}
    \label{eq:cutoff-KK}\Lambda\overset{!}{>}m_1\times \left(\frac{m_N}{m_1}\right)^{3/2}=m_1\times \left(\frac{6~\text{TeV}}{m_1}\right)^{3/2}\,.
\end{equation}
Notice how for $m_1=10$~GeV, this already implies a strong lower bound on the interaction scale $\Lambda\gtrsim 150$~TeV.

\bigskip
In scenarios with $d>1$ compact extra dimensions, KK-modes are characterized by a $d$-dimensional KK-vector composed of integers. The detailed mass spectrum and couplings depend on the specific model. If the extra dimensions share similar properties, such as in models with universal extra dimensions~\cite{Arkani-Hamed:1998jmv}, the KK tower can exhibit a much denser spectrum of graviton masses, and in extreme cases even (quasi-)degenerate modes. To illustrate the impact of KK multiplicities, we consider the benchmark spectrum 
\begin{align}
    &m_{\Vec{n}_d} = |\Vec{n}_d|\times m_1\,, &c_{\Vec{n}_d}=1\,.
\end{align}
where $\Vec{n}_d=(n_1,\dots, n_d)$ is a vector of dimension $d$ whose entries are non-negative integers.  It should be stressed that this is only an illustrative example: some extra dimensions could have different sizes, non-zero curvature or a non-trivial topology, or different KK number structure due to space-time symmetries. For example, allowing negative integer components of $\Vec{n}_d$ would further increase the density of the KK spectrum.

\section{Methodology}
\label{sec:method}

We consider the invariant mass distribution for the production of a pair of leptons (electrons or muons) or a pair of photons at the LHC at hadronic center-of-mass energy $\sqrt{s}=13$ TeV, $\frac{\d\sigma}{\d m_{\ell\ell,\gamma\gamma}}$. The measured yields in each bin of the differential distribution, as well as the expected number of events from the Standard Model background, are obtained from the ATLAS and CMS experiments.
We then compute the signal due to one or more massive gravitons decaying into a pair of leptons or photons, and place constraints on the parameter space of the interaction Lagrangian in~\eqref{eq:interactionLagrangian}, namely on $(m_1,\Lambda)$.
In the following, we illustrate how we obtain the experimental data and the relevant SM background, compute the graviton signal, and extract our constraints. The code used in this work is publicly available for reproducibility~\cite{GravBounds}.
\subsection{Data and background}\label{sec:data}
We consider dilepton production data measured during Run II by ATLAS and CMS~\cite{ATLAS:2019erb,CMS:2021ctt}, as well as diphoton production data from ATLAS~\cite{ATLAS:2021uiz}. Analogous diphoton data from the CMS experiment exists (see e.g. Ref.~\cite{CMS:2024nht}) but could not be implemented in our analysis due to the lack of a public release of the mass resolution model. In Refs.~\cite{ATLAS:2019erb,CMS:2021ctt,ATLAS:2021uiz}, the measured invariant mass distributions of the dilepton (diphoton) system are compared to the expected SM results to search for a single resonant signal from a selection of BSM models.
For the ATLAS dilepton (diphoton) analysis, we consider the mass range $225, 5806.19$ GeV ($150, 2400$ GeV), respectively. Instead, from the CMS dilepton analysis we use data in the invariant mass range of $150, 3570$ GeV for dielectron and $150,2908.1$ GeV for dimuons.
References~\cite{ATLAS:2019erb,CMS:2021ctt,ATLAS:2021uiz} also provide functional forms for the SM background of $\frac{\d \sigma}{\d m_{\ell \ell,\gamma \gamma}}$. We adopt these to compute the SM background for our analysis in place of generating our own SM samples.

\subsection{Signal generation}

The signal is computed with an in-house code~\cite{GravBounds}.
We consider the production of a graviton $G$ with mass $m_G$ and width $\Gamma_G$ in proton-proton collisions at centre of mass energy $\sqrt{s}$. The graviton then decays into a generic two-body final state $F$. The decay is taken in the narrow-width approximation~(NWA), where the graviton's Breit-Wigner distribution is replaced by a delta-function:
\begin{equation}
\label{NWAppr}
    \frac{1}{\pi}\frac{m_G\Gamma_G}{(m_F^2-m_G^2)^2+m_G^2\Gamma_G^2} \quad \longrightarrow \quad \delta(m_F^2-m_G^2) \ ,
\end{equation}
with $m_F^2$ being the squared invariant mass of the graviton's decay products.
Both production and decay of the graviton are treated at Born level.
Under these assumptions, the differential cross section in the invariant mass of the final state reads 
\begin{eqnarray}
\label{eq:crossPPNW}
        \frac{\d \sigma}{\d m_F}  &=& \frac{16\pi^2}{s m_G} \Gamma_G   B_F \sum\limits_{i,j} g_{ij}   B_{I,ij}  \mathcal{L}_{ij} \delta(m_F - m_G),
\end{eqnarray}
where the sum runs over the pairs of initial-state partons $i$ and $j$, $B_{I,ij}$ and $B_F$ are the production and decay branching ratios, and $g_{ij}$ encodes the spin- and colour-average of the initial state and the graviton multiplicity. 
\begin{table}
    \centering
    \renewcommand{\arraystretch}{1.3} 
    \setlength{\tabcolsep}{12pt} 
    \resizebox{\textwidth}{!}{
    \begin{tabular}{|c|c|c|}
        \hline
        \textbf{Dataset} & \textbf{$p_T$ cuts} & \textbf{Rapidity cuts} \\
        \hline\hline
        ATLAS $e^- e^+$ & $p_{T} > 30\, \text{GeV}$ & $|y| < 1.37 \lor 1.52 < |y| < 2.47$ \\
        ATLAS $\mu^- \mu^+$ & $p_{T} > 30\, \text{GeV}$ & $|y| < 2.50$ \\
        ATLAS $\gamma \gamma$ & $p_{T} > \max(0.3\, m_G, 25\, \text{GeV})$ & $|y| < 1.37 \lor 1.52 < |y| < 2.37$ \\
        \hline
        CMS (barrel-barrel) $e^- e^+$ & $p_{T} > 35\, \text{GeV}$ & $|y| < 1.44$ \\
        CMS (barrel-barrel) $\mu^- \mu^+$ & $p_{T} > 53\, \text{GeV}$ & $|y| < 1.44$ \\
        \multirow{2}{*}{CMS (barrel-endcap) $\ell^- \ell^+$} & \multirow{2}{*}{same as barrel-only} & one lepton in $1.57 < |y| < 2.40$  \\
        & &  and one in $|y| < 1.44$  \\
        \hline
    \end{tabular}
    }
    \caption{Fiducial cuts for dilepton and digamma searches from ATLAS and CMS, see Refs.~\cite{ATLAS:2019erb,ATLAS:2021uiz,CMS:2021ctt}. The $p_T$ and $y$ cuts are applied to individual leptons (photons).}
    \label{tab:experiment_cuts}
\end{table}
Finally, $\mathcal{L}_{ij}$ is the parton luminosity, given by:
\begin{equation}
\label{eq:lumi}
\mathcal{L}_{ij}=\int\limits_{m_G^2/s}^1 d\xi
        \frac{f_i(\xi)}{\xi}f_j\left(\frac{m_G^2}{s \xi}\right) \vartheta_F \,,
\end{equation}
with $f_{i}$ representing the proton Parton Distribution Function (PDF) associated to parton $i$ and $\vartheta_F$ is an acceptance function enforcing the experimental fiducial cuts, which we impose in our implementation and are summarised in Table~\ref{tab:experiment_cuts} for the considered datasets.
We use the \texttt{NNPDF3.1\_lo\_as\_0118} PDF set, and we assessed that PDF-member variations, as well as different PDF choices, lead to small effects on the final bounds. 
As a sanity check of our implementation, we verify that our results for the inclusive cross section are in full agreement with \texttt{PYTHIA 8.3}~\cite{Bijnens:2001gh,Bierlich:2022pfr}. We also checked that our signal evaluation is in good agreement with the results reported
in Refs.~\cite{ATLAS:2019erb,CMS:2021ctt,ATLAS:2021uiz} obtained for specific values of $\Gamma_G/m_G$. 

The decay widths of a massive graviton into SM final states for arbitrary values of the mass can be found in App.~\ref{app:width-SM}. For a graviton much heavier than its decay products, the expressions simplify, and the approximate branching ratios are given in Tab.~\ref{tab:branching}.
\begin{table}[]
    \centering
    \begin{tabular}{|c|ccccccc|}
    \hline
       Final State  & $gg$ &$\gamma\gamma$ &$\sum_q\Bar{q}q$ &$\sum_\ell\Bar{\ell}\ell$ &$W^+W^-$& $ZZ$ & $hh$\\
       \hline
       Branching Ratio [\%]  & $32.9$ & $4.1$& $37.0$ & $12.3$  & $8.9$& $4.5$& $0.3$\\
    \hline
    \end{tabular}
    \caption{Approximate tree-level branching ratios of a graviton into SM two-body final states assuming it to be heavier than the top-quark pair production threshold, $m_G\gg 2m_t$. Each decay channel into a pair of colour-singlet fermions contributes about $2\%$. }
    \label{tab:branching}
\end{table}
In the same limit, the total graviton's width into SM final states reads
\begin{equation}
    \label{eq:SM-width}\Gamma_{G} \approx  0.097\frac{m_G^3}{\Lambda^2}\equiv \eta_G \frac{m_G^3}{\Lambda^2}\,.
\end{equation}
Some consistency constraints descend implicitly from our assumptions and reduce the parameter space we can study. First, if the graviton's width is to be narrow $\Gamma_G\ll m_G$, then 
\begin{equation}
    \frac{m_G}{\Lambda}\ll \eta_G^{\frac{1}{2}}\approx 0.3\,.
\end{equation}
As it will turn out, this condition is excellently satisfied in the entire parameter space here considered, with the sole exception of the right tail of the mass spectrum, where $m_G\gtrsim 4$~TeV.
Secondly, for the graviton decay to be prompt, the ratio $m_G/\Lambda$ cannot be arbitrarily small. Assuming the allowed displacement to be smaller than a millimetre, for a relativistic graviton with energy $E$, one finds 
\begin{equation}
  \label{eq:prompt}
  \frac{m_G}{\Lambda}\gtrsim 2\times 10^{-7}\times \left(\frac{200~\text{GeV}}{m_G}\right)\left(\frac{E}{1~\text{TeV}}\right)^{1/2}\,.
\end{equation}
As we will see, the probed region of the parameter space satisfies this condition.

To realistically simulate the contribution of the New Physics (NP) signal, we take into account the detector efficiency and its mass resolution in the reconstruction of different decay products of the graviton. To implement these final-state dependent effects, we rely on the efficiency and mass-resolution model parameters reported by the experiments~\cite{ATLAS:2019erb,CMS:2021ctt,ATLAS:2021uiz}. The limited mass resolution of the detector smears the invariant mass delta-function in Eq.~\eqref{NWAppr} into a finite-width peaked distribution. This is necessary to obtain realistic bounds, which would otherwise be largely overestimated. 
For the ATLAS diphoton sample, we apply a flat $70$\% efficiency factor, following Run II calibration from Ref.~\cite{ATLAS:2019qmc}.
The CMS dilepton analysis~\cite{CMS:2021ctt} did not come with a public mass-resolution model. Upon discussion with the analysis authors, we adopted the peak parametrisation from the 2018 dataset as an approximate stand-in.
Similarly, since we could not recover the appropriate mass resolution model for the data in Ref.~\cite{CMS:2024nht}, we omitted diphoton data from CMS from our analysis.

\subsection{Likelihood profiling} \label{ssec:nll-pl}

To derive the constraints, we employ a frequentist approach (see e.g. Ref.~\cite{Junk:1999kv} for a review) and approximate the likelihood function for each bin as a Poissonian distribution. 
For each bin $i$, the experimentally observed counts are denoted as $n_i$, while the expected yield is
\begin{equation}
\lambda_i(\Lambda,\alpha,\theta)=b_i(\alpha)+ \frac{s_i(\theta)}{\Lambda^2},    \label{eq:poisson-mean}
\end{equation}
where $b_i$ is the SM background contribution, modelled as a function of a set of parameters $\alpha$ (see Sec.~\ref{sec:data}), and $s_i$ is the NP graviton signal stripped of its overall coupling $\frac{1}{\Lambda^2}$.  
The symbol $\theta$ denotes a collection of nuisance parameters that quantify systematic uncertainties affecting the signal. These are applied as linear deformations, \mbox{$s_i(\theta) \to s_i (1 + \delta_{i} \cdot \theta)$}, with sizes $\delta_i$ extracted from the respective experimental analyses and $\theta$ following a normal Gaussian distribution. The full list and sizes of nuisance parameters are in App.~\ref{app:systematics}.
Given $N$ bins, the Poisson negative log-likelihood $\mathcal{L}$ (NLL) then reads
\begin{align}
    &- \log \mathcal{L}(n,\lambda)= \sum_i \left[\lambda_i-n_i\log \lambda_i\right]
+\frac12\sum_k \theta_k^2\,,
    \label{eq:likelihood}
\end{align}
up to an additive constant. The second term parametrizes the normal Gaussian penalty introduced by the nuisance parameters.

We define our null hypothesis by setting $ s_i \equiv 0$ and minimizing the NLL as a function of the experimental background model parameters.
\begin{equation}
    \alpha_0 = \arg\min_{\alpha}\bigl[-\log \mathcal{L}(n,\lambda)\bigr],
\end{equation}
this yields a reference value $\mathcal{L}_0$.
Next, for any value of $\Lambda$, we compute the same estimation for the signal plus background hypothesis, this time including nuisance parameters in the fit. We obtain the working point,
\begin{equation}
    (\hat\theta(\Lambda),\hat\alpha(\Lambda)) = \arg\min_{\theta,\alpha}\bigl[-\log \mathcal{L}(n, \lambda)\bigr],
\end{equation}
which, in turn, allows us to build the profiling statistic
\begin{equation}
    q(\Lambda)=2\left[-\log \mathcal{L}\bigl( n, \lambda(\Lambda, \hat{\alpha}, \hat{\theta}) \bigr) + \log \mathcal{L}_0\right],
\end{equation}
that quantifies how much worse the fit becomes when a signal scaled by $\Lambda$ is present.
To perform our limit extraction of the coupling, we employ Wilks' theorem~\cite{Wilks:1938dza}, and identify $q(\Lambda)$ as a $\chi^2$ variable with one degree of freedom. We scan across values of $\Lambda$ and identify the bounds as the lowest value for lambda $\Lambda_{95\%}$ ($\Lambda_{99\%}$) such that $q = 3.84$ ($q = 6.64$).

\section{Analysis}
\label{sec:analysis}

\subsection{Single Graviton}
\label{sec:single}

\begin{figure}
    \centering
    \includegraphics[width=\linewidth]{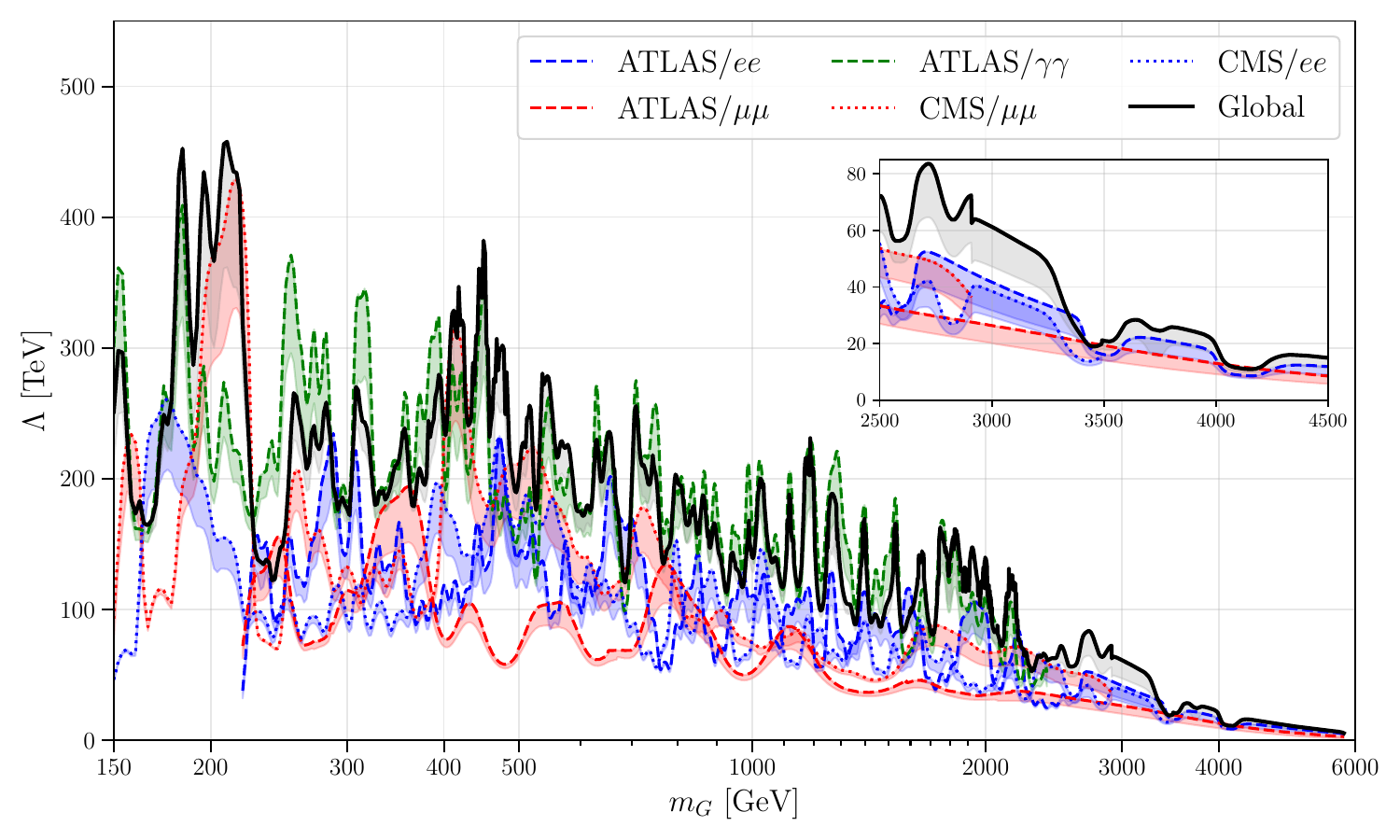}
    \caption{Lower bounds at the $95\%$ and $99\%$~C.L., represented by the upper and lower edges of the shaded bands, on the interaction scale $\Lambda$ for a single graviton employing data from ATLAS~(dashed) and CMS~(dotted) including the diphoton~(green), dielectron~(blue) and dimuon~(red) final states. The comprehensive constraint derived from the combination of the different dataset likelihoods is shown with a solid black line.
    }
    \label{fig:single-constraints}
\end{figure}

We begin by reviewing the single graviton constraints and extending them by combining the likelihood of the experiments. The resulting $95\%$ and  $99\%$ C.L. lower bounds on the interaction scale $\Lambda$ for a single graviton are shown in Fig.~\ref{fig:single-constraints}. We show individual shaded areas corresponding to limits obtained from ATLAS and CMS data in the diphoton, dielectron, and dimuon final states. We also include the combined constraint obtained by multiplying the likelihoods of all available channels from both experiments. Possible correlations between the datasets are not taken into account, as no correlation matrix for the analyses is publicly available to the best of our knowledge. 

For masses smaller than $m_G\lesssim 2.5$~TeV down to $500$~GeV, the ATLAS diphoton data dominate the scenario and are able to set constraints at the level of $\Lambda\sim 150-300$~TeV. For smaller masses, the bounds get even stronger, reaching spikes of $\sim500$~TeV. As the mass grows, dilepton final states take over and the lower bound decreases from $\Lambda\sim 60$~TeV to $\Lambda\sim10$ TeV at $m_G\sim 6$~TeV. The combined constraints, shown with a black line in the plot, overall follow the pattern dictated by the two aforementioned regimes. A few exceptions arise, where the combined Poissonian statistics of the data can either increase or lower the bound by up to $30\%$. These two cases are well visible in the ranges $180-250$~GeV, and $250-350$~GeV, respectively, where the constraints gain or lose about $100$~TeV when the datasets are combined.

\bigskip
Let us compare the results to previous findings. Experimental searches typically set constraints focusing only on the lightest KK mode in the parameter space $(m_1,k/\overline{M}_P)$, where $k$ is the curvature of the RS model, and $\overline{M}_P$ is the reduced Planck mass. 
The reason for constraining such a quantity is convenience. Indeed, in the NWA, the NP signal amplitude is entirely encoded in the ratio of the graviton width over its mass (cfr.~\eqref {eq:crossPPNW})
\begin{equation}
    \frac{\Gamma_G}{m_G}
    \approx 1.42 \left(\frac{k}{\overline{M}_P}\right)^2\,.
\end{equation}
A value of $k/\overline{M}_P$ too small makes the graviton long-lived, thus invalidating our analysis. By employing the lower limit on $\Lambda/m_G$ of Eq.~\eqref{eq:prompt}, for a graviton with energy $E$, this translates into
\begin{equation}
  \label{eq:koverM-lower}
  \frac{k}{\overline{M}_P}\gtrsim 6\times 10^{-8}\times\left(\frac{200~\text{GeV}}{m_G}\right)\left(\frac{E}{1~\text{TeV}}\right)^{1/2}\,.
\end{equation}
This value is orders of magnitude smaller than what is typically analysed by the experimental collaborations, which reach values of only $k/\overline{M}_P=0.01$~\cite{ATLAS:2021uiz,CMS:2024nht}.

\begin{figure}
    \centering
    \includegraphics[width=\linewidth]{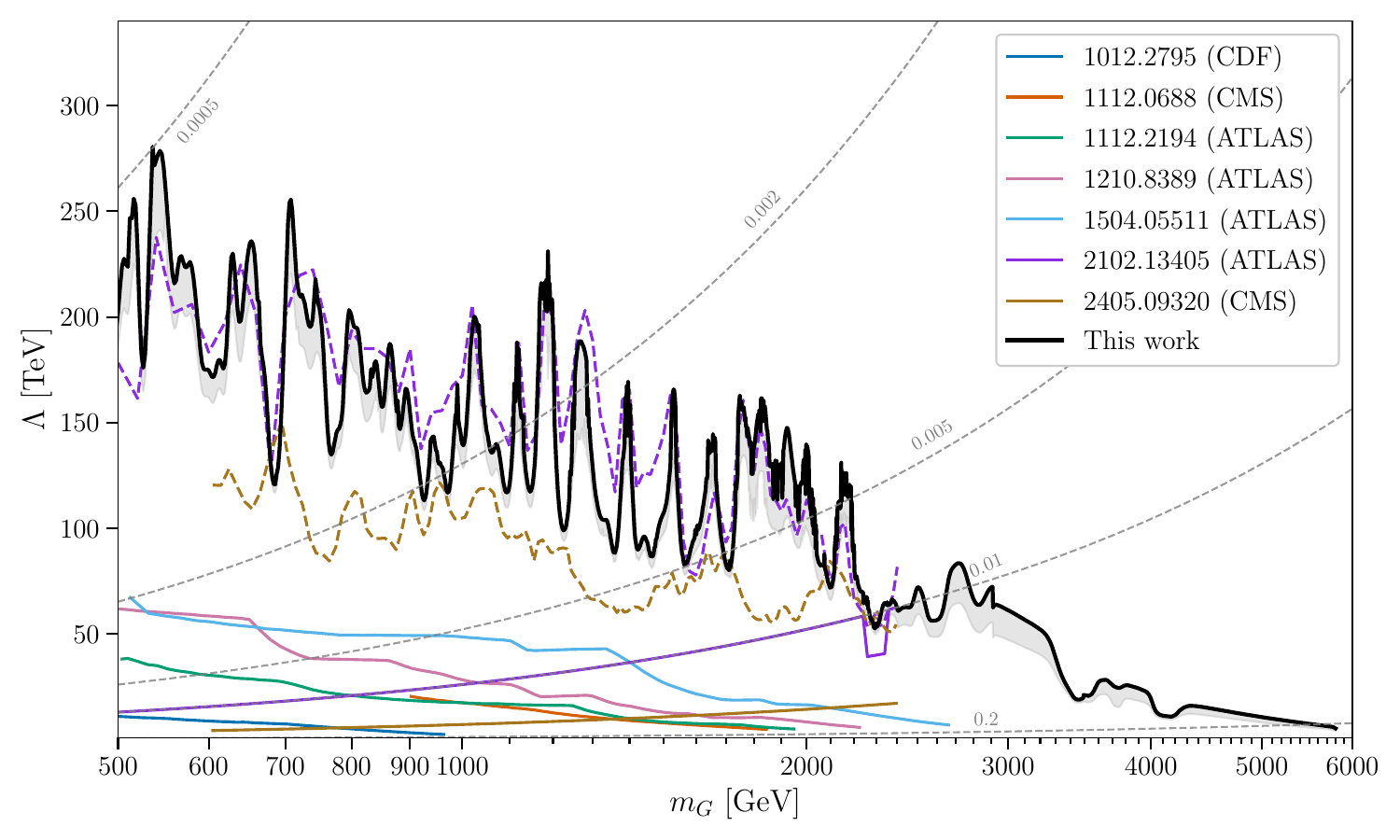}
    \caption{Comparison between the lower bounds on $\Lambda$ derived in this work at $95\%$ and $99\%$~C.L. (solid black and grey shaded area, respectively) with previous literature (different colours). We report in the legend the corresponding arXiv number. Contours of different values of $k/\overline{M}_P$ are shown with dashed gray lines.}
    \label{fig:bounds-comparison}
\end{figure}

Employing the expression of the graviton width of Eq.~\eqref{eq:SM-width}, this translates into an exclusion of the pairs $(m_G,m_G/(\gamma_1\Lambda))$, where $\gamma_1\approx 3.83$ is the first zero of the Bessel-$J_1$ function. 
In the $(m_G,\Lambda)$ planes, such exclusion curves translate into triangular-shaped exclusion regions. 
We show in Fig.~\ref{fig:bounds-comparison} the comparison between the our total constraint of Fig.~\ref{fig:single-constraints} and the constraints derived from a representative selection of previous studies~\cite{CDF:2010muc,CMS:2011bsw,ATLAS:2011ab,ATLAS:2012hvw,ATLAS:2015shg,CMS:2016kgr,ATLAS:2021uiz,CMS:2024nht}.
As can be seen, there has been remarkable progress achieved over the past 15 years in setting increasingly stringent constraints on $\Lambda$. However, the experimental focus on a limited range of $k/\overline{M}_P$ values effectively restricts the explored region of the $(m_G, \Lambda)$ parameter space. Given the lower bound on $k/\overline{M}_P$ of Eq.~\eqref{eq:koverM-lower}, there is no fundamental reason to exclude smaller values from the analysis.
We suspect that this region may have been overlooked due to a theoretical bias, since scenarios with $k/\overline{M}_P \ll 1$ are often regarded as ``unnatural'' on the model-building side. For the sake of comparison, we extend the data to the region of smaller $k/\overline{M}_P$ by employing the upper bound on the cross-section for the smallest available $k/\overline{M}_P$.

In App.~\ref{app:comparison-ATLAS} and App.~\ref{app:comparison-CMS} we present the comparison between the single-graviton constrains we obtain in this work with the ones reported or extracted from the ATLAS and CMS analysis respectively. We found good agreement across all considered final states.

\subsection{Kaluza-Klein Gravitons in One Extra Dimension}
\label{sec:multiple}

In this Section, we go beyond the single-graviton approximation considered so far and derive constraints including the entire tower of KK modes as outlined in Sec.~\ref{sec:models}. 

Before proceeding with the analysis, it is worth noticing that,  compared to the single-graviton case, there is a subtlety. While the lightest graviton can only decay into SM particles, heavier KK-gravitons can also decay into lighter ones. Given the large number of SM final states, this is not generically a problem, but if the KK-number is large enough, gravitons' final states could dominate. If this happens, the branching ratio into SM gets reduced, and the bounds weaken. In the most extreme case, heavy KK-gravitons could decay entirely into lighter gravitons, thus vanishing attempts to set constraints from diphotons/dileptons final states. 
However, this turns out not to be the case due to the non-trivial structure of couplings and masses of the KK gravitons. For example, within the RS model discussed in~App.~\ref{app:comparison-widths}, the branching ratios into KK final states reach at most a $\sim (1-1.5)\%$ for $m_1\in(5~\text{GeV},1~\text{TeV})$. For all masses, they actually decrease as $m_n$ increases. We will therefore neglect such final states and focus only on SM ones in the following analysis.

The presence of multiple resonances will affect multiple bins simultaneously. The smaller $m_1$, the denser the number of resonances per bin.
Naively, given the data considered in this work, one could argue that more resonances imply a stricter constraint. However, this is not the case. There are two reasons for this, one theoretical and one experimental.
About the former one, as it was argued in Sec.~\ref{sec:models}, lowering the mass of the lightest state also lowers the cutoff of the theory, thus undermining the theoretical reliability of the computation. If $m_1$ gets too small, then the theoretical lower bound on $\Lambda$ becomes so large that experimental data cannot compete. From that side, the lowest mass we could consider is about $m_1=10$~GeV, where the cutoff condition of Eq.~\eqref{eq:cutoff-KK} translates to
\begin{equation}
    \Lambda\gtrsim 150~\text{TeV}\times \left(\frac{10~\text{GeV}}{m_1}\right)^{1/2}\,.
\end{equation}
For $m_1=1$~GeV, the lower value grows to $\sim 450$~TeV. As it will turn out, this condition is close to the value of the scale that can actually be probed, thus beginning to undermine the reliability of the results on the theoretical side.
\begin{figure}
    \centering
    \includegraphics[width=\linewidth]{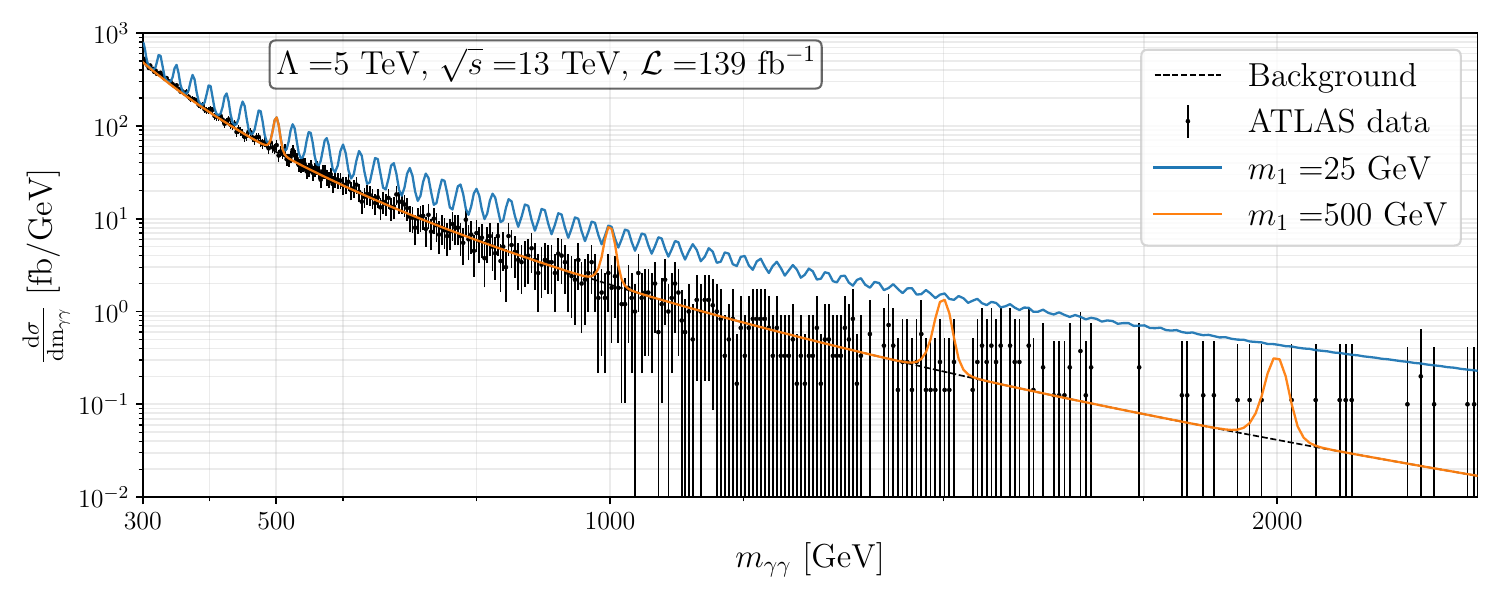}
    \caption{Visualisation of the KK gravitons' resonances in the region $m_{\gamma \gamma} \in \left(300, 2300\right)$~GeV over the SM background~(dashed black line) and data in the ATLAS diphoton channel~\cite{ATLAS:2021uiz}. For illustrative purposes, we set $\Lambda=5$~TeV and $m_1=\{25,500\}$~GeV in blue and orange, respectively.}
    \label{fig:resonances}
\end{figure}

Turning to the experimental one, if $m_1$ decreases too much, the detector resolution smears the resonances enough that the peak structure effectively converts to a continuum. At that point, the background function can absorb the NP signal, thus losing sensitivity to $\Lambda$.
For the sake of illustration, we exemplify such a phenomenon in Fig.~\ref{fig:resonances} for a representative set of benchmark masses, $m_1=\{25,500\}$~GeV and $\Lambda=5$~TeV. We show the signal of the entire KK tower over the background for the ATLAS diphoton channel. For $m_1=500$~GeV, the NP signal is composed of a structure of well-separated and recognisable resonances. As $m_1$ decreases, the resonances begin to overlap, and only a high-frequency series of peaks remains visible in the low $m_{\gamma\gamma}$ region. In the high $m_{\gamma\gamma}$ tail, the smearing of the detector destroys the structure of peaks, leading to a continuum, which could be fitted by a background function. When this happens, the constraining power rapidly diminishes, suggesting that alternative strategies are needed, such as restricting the dataset to lower masses, or constraining the background function. While this is per se an interesting question, it goes beyond the scope of this work.
In order to avoid this issue altogether and ensure the meaningfulness of the constraint employing the given background function, we impose a lower cut on $m_1$ for each data set. The cut is chosen differently for each data set, ensuring that graviton resonances remain distinguishable from a continuum. In practice, since the mass resolution model width generally increases with reconstructed $m_{\ell\ell,\gamma \gamma}$, we adopt a conservative strategy requiring
\begin{equation}
    m_1 \geq 2\sigma_{\mathrm{res.}}( M ) \,,
\end{equation}
where $\sigma_{\rm res.}$ is the standard deviation of the mass-resolution model at mass $M$, which we pick at the rightmost edge of the last populated bin of each dataset.
We choose $(M,\sigma_{\mathrm{res.}})$  to be $(2640\, \mathrm{GeV}, 100\, \mathrm{GeV})$,  $(2470\, \mathrm{GeV},  340\,\mathrm{GeV})$, and $(2400\,\mathrm{GeV}, 20\,\mathrm{GeV})$ for ATLAS $ee,\mu\mu$, and $\gamma \gamma$, respectively.
For CMS dielectrons and dimuons, we set instead $(3570\,\mathrm{GeV}, 200\,\mathrm{GeV})$ and \mbox{$(2908\,\mathrm{GeV}, 380\,\mathrm{GeV})$}.

\begin{figure}
    \centering
    \includegraphics[width=\linewidth]{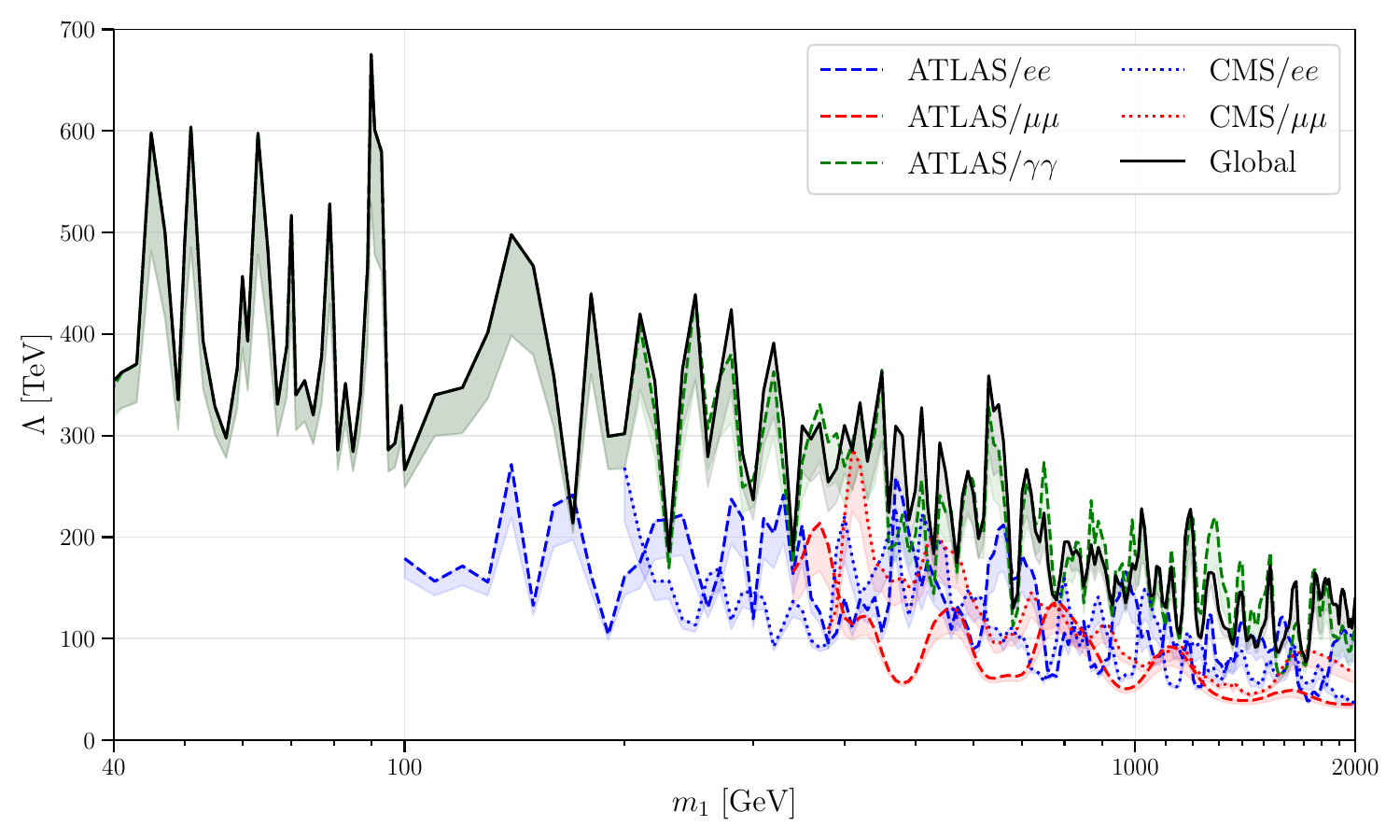}
    \caption{Lower bound at the $95\%$~C.L. on the interaction scale $\Lambda$ including the whole tower of gravitons employing data from ATLAS~(dashed) and CMS~(dotted) including the diphoton~(green), dielectron~(blue) and dimuon~(red) final states. The comprehensive constraint derived from the combination of the different dataset likelihoods is shown with a solid black line.}
    \label{fig:KK-constraints}
\end{figure}

\bigskip
We set a lower bound on the extra-dimensional scale $\Lambda$ as a function of the lightest mass $m_1$ of the tower. We scan $m_1$ from $40$~GeV to $2$~TeV. The lower bound is set by the requirement that the resonances at large masses do not merge into an effective continuum indistinguishable from the background. The upper bound instead is chosen purely for convenience: above $2$~TeV, the heavier modes of the tower fall in a region of the data with very little constraining power, so the resulting bounds are essentially identical to those derived for the single graviton in Sec.~\ref{sec:single}.
The result can be seen in Fig.~\ref{fig:KK-constraints}.
\begin{figure}
    \centering
    \includegraphics[width=\linewidth]{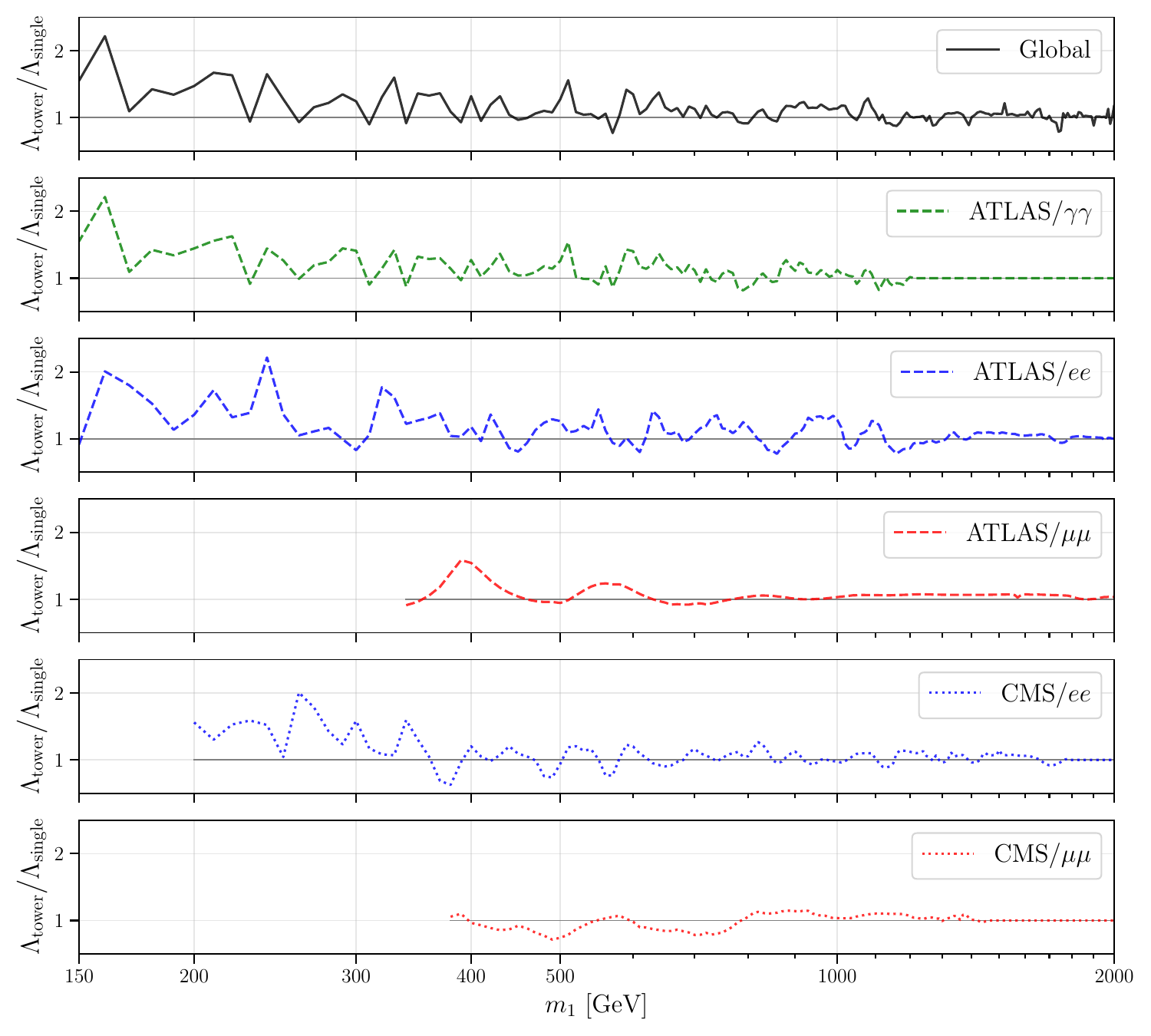}
    \caption{Ratio of the bounds at the $95\%$~C.L. derived including the whole tower of gravitons ($\Lambda_\text{tower}$) with respect to the one with a single graviton ($\Lambda_\text{single}$) for different datasets of ATLAS and CMS.}
    \label{fig:KK-constraints-ratio}
\end{figure}
The tower plays a crucial role in the low-mass region, below $150$~GeV, which would otherwise be unconstrained by the dataset analysed here. Its inclusion allows us to set a lower bound in that region, ranging from $400$ to $800$~TeV. At larger masses, the gain relative to the single-graviton case varies across the spectrum, reaching up to a factor of $2$ and typically falling between $1$ and $1.5$. Fig.~\ref{fig:KK-constraints-ratio} shows the ratio of the constraint from each dataset to the single-graviton case, providing a clearer view of these gains.

As can be seen, aside from the expected enhancement of the constraints, for some mass values the constraint becomes weaker rather than stronger. This is due to statistical fluctuations of the data: even though the whole tower is present, typically only a handful of resonances actually drive the constraint. If, in the single-graviton case, a single resonance happens to fall in a downward fluctuation of the data, it produces a very strong constraint; once an extra resonance is added, it is likely to fall in a less constraining region, thus weakening the bound. In extreme cases, the extra resonance may fall in a bin where the data fluctuate upward, making the overall bound noticeably smaller, by as much as $20\%$ in some cases.

Our results show that including the full tower of massive gravitons leads to non-trivial modifications of the constraints on the interaction scale $\Lambda$, and extends the bound-placing procedure to lower graviton masses, thereby fully exploiting the constraining power of the data. We therefore advocate for the inclusion of the complete spectrum in future searches for spin-2 massive resonances, in order to derive constraints that are appropriate to specific models.

\subsection{Kaluza-Klein Gravitons in Multiple Extra Dimensions}
\label{sec:multiple-d}

We now turn to the case of multiple extra dimensions. 
As outlined in Sec.~\ref{sec:models}, we choose the spectrum define by
\begin{align}
    \label{eq:multiple-spec}(m_{|\Vec{n}_d|}, \Lambda_n)=(|\Vec{n}_d|m_1, \Lambda)\,,
\end{align}
where $\Vec{n}_d$ is a vector composed of non-negative integers. The larger multiplicity of modes can significantly increase the NP signal in two ways. If the modes are strictly degenerate, then the same resonance gets amplified. For example, in the case of Eq.~\eqref{eq:multiple-spec}, in $d=2$, modes with $(1,0)$ and $(0,1)$ have the same mass, and the resonance gets larger by a factor of $2$. In $d$-dimensions, this gets replaced by a factor of $d$. The same fate applies to resonances associated with more complicated $\Vec{n}_d$, and the multiplicity rapidly grows proportionally to $|n_d|^{d-2}$. Secondly, new resonances appear. For example, in $d=2$, a new mode with mass $\sqrt{2}m_1$ appears, thus further increasing the number of bins affected. In such a scenario, the cutoff scale is lower compared to the single extra-dimensional case.

\begin{figure}
    \centering
    \includegraphics[width=\linewidth]{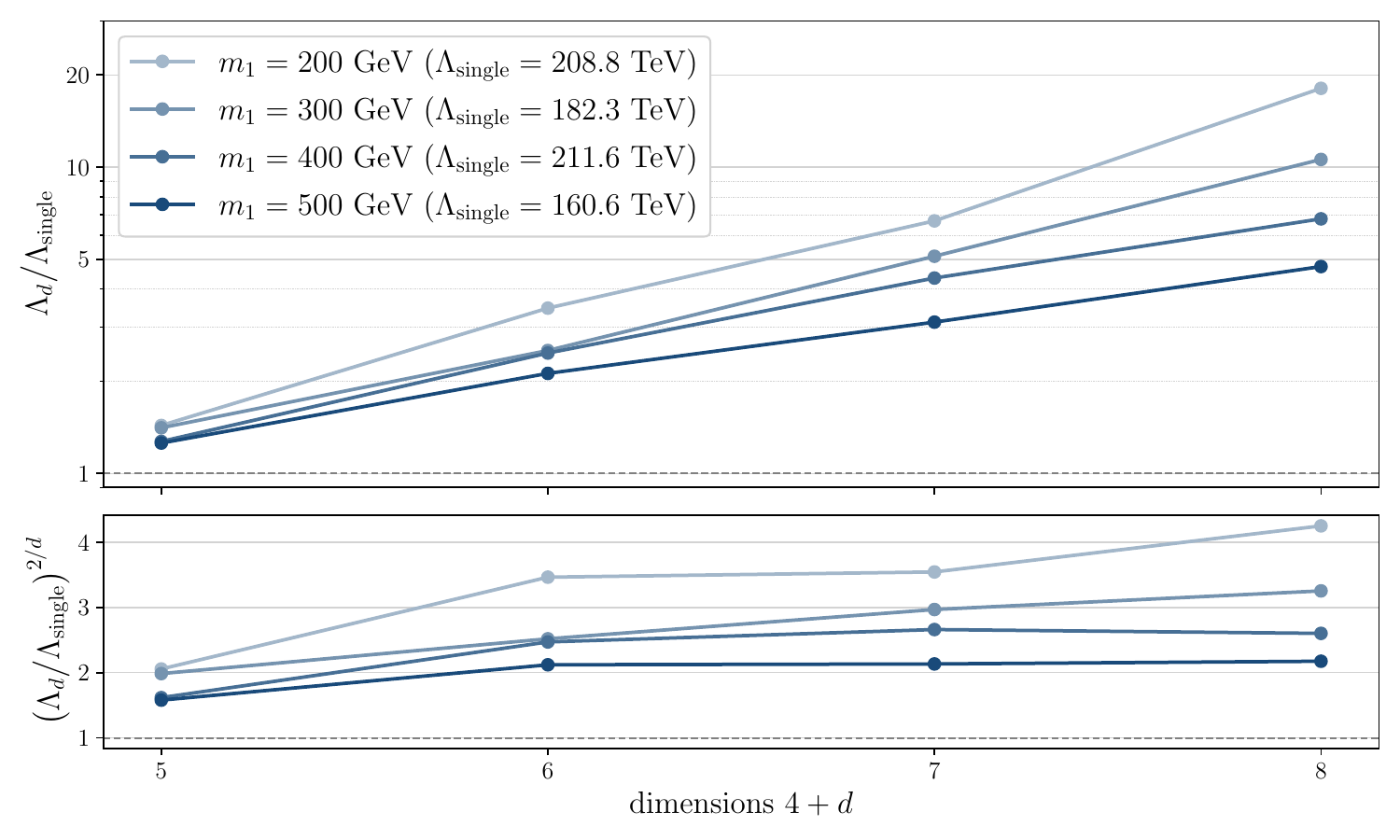}
    \caption{Constraints on $\Lambda$ from towers of gravitons in multiple dimensions. The upper panel shows the ratio to the constraint obtained in the single-resonance case, while in the lower panel we assess the scaling of the bounds as the number of extra dimensions increases.
    }
    \label{fig:KK-constraints-over-D}
\end{figure}

Given the large model dependence that enters in scenarios with more than an extra dimension, we focus here on the gain that such models would most likely produce. To this end, we consider the most stringent dataset, ATLAS-$\gamma\gamma$, select a few benchmarks for the lightest mode, $m_1\in\{200,300,400,500\}$~GeV, and plot the ratio with respect to the single graviton case for different numbers of extra dimensions $d=1,2,3,4$. The result can be seen in the upper panel of Fig.~\ref{fig:KK-constraints-over-D}. As expected, as $d$ grows, the bound gets stronger. An analytical estimation of the growth is challenging, but we can get a rough idea using two different arguments. The first one: let us consider the constraint derived from the total mass-integrated yield, summing over the contributions of all the resonant states. We define by $N$ the KK number associated to the heaviest state, which is nevertheless light enough to contribute significantly to the data. Then, the number of lighter states that contribute grows as $N^d$, and the constraint would scale as
\begin{eqnarray}
    \frac{\Lambda_d}{\Lambda_\text{single}}\propto C^{d/2},
\end{eqnarray}
where $C$ is a constant and the scaling holds up to weaker $d$-dependent terms.
We could also take another path: if one assumes that each bin in the differential distribution is equally relevant, one can approximate the sum to an integral, and find
\begin{equation}
    \chi^2\sim \int\limits_1^N dn \left(\frac{n^2}{\Lambda^2}\times n^{d-2}\right)^2\,,
\end{equation}
where the $n^2$ factor accounts for the effective $m_n^2$ dependence of the signal of Eq.~\eqref{eq:crossPPNW}, and $n^{d-2}$ for $d\geq 2$ accounts for the multiplicity of states such that $|\Vec{n}_d|^2=n^2$ and it is equal to $1$ for $d=1$. 
All in all, this leads to
\begin{equation}
   \frac{\Lambda_d}{\Lambda_\text{single}} \propto \begin{cases}
       (C)^{5/4} & d=1\,,\\
       (C)^{d/2+1/4} & d\geq 2\,,
   \end{cases}
\end{equation}
again up to terms which grow slower at large $d$. 
All in all, at large $d$ one expects
\begin{eqnarray}
    \log\left(\frac{\Lambda_d}{\Lambda_\text{single}}\right)\propto\frac{d}{2}+\dots\,.
\end{eqnarray}
Such scaling is shown in the bottom panel of Fig.~\ref{fig:KK-constraints-over-D}. While it cannot capture the exact dependence on the data, nor the exact prefactor, the curve is reasonably flat, thus supporting our argument.

\section{Conclusions}
\label{sec:conclusions}

Extra-dimensional theories of gravity are among the best-motivated frameworks for physics beyond the Standard Model, and their most robust prediction is the existence of a tower of massive spin-2 resonances. While experimental searches for such states have traditionally focused on the lightest mode alone, the full tower can play a significant role in shaping the accessible parameter space, both by strengthening bounds where multiple resonances contribute and by extending sensitivity to mass regions that a single-resonance treatment would leave unconstrained.

In this work, we revisited the LHC constraints on resonant massive gravitons using dilepton and diphoton data from ATLAS and CMS. We first updated the single-graviton bounds on the interaction scale $\Lambda$, and validated our procedure against the results of the experimental collaborations in Sec.~\ref{sec:single}. We then extended the analysis to include the entire Kaluza-Klein tower, considering both a single extra dimension in the Randall-Sundrum setup in Sec.~\ref{sec:multiple}, and generalisations to multiple extra dimensions with denser spectra in Sec.~\ref{sec:multiple-d}.

We found that the inclusion of the full tower can significantly alter the constraint, in some cases strengthening it by a factor of $2$ compared to the single-graviton treatment. Interestingly and somewhat counter-intuitively, occasional statistical fluctuations can instead weaken it: a single resonance landing on a downward fluctuation of the data can produce an especially strong bound, and adding neighbouring resonances from the tower can dilute rather than reinforce this effect. The impact of the tower becomes crucial at lower masses, below the electroweak scale, where the density of resonances is large enough that the tower alone allows us to set competitive constraints. It is worth pointing out that those masses cannot be probed with the dataset used in this study in the single resonance scenario. In the case of multiple extra dimensions, the growth of the bound with the number of extra dimensions $d$ follows an approximate power law in $d/2$, driven by the increasing multiplicity and density of degenerate and near-degenerate modes.
The toolset used to produce these results is publicly available~\cite{GravBounds}.

Aside from the direct application in phenomenology, we hope that this analysis can serve as a template for experimental collaborations to systematically incorporate the full graviton tower in future resonance searches, rather than restricting the interpretation to the lightest mode alone. A natural extension of this work is the study of projected sensitivities for the High-Luminosity phase of the LHC. We expect the derived constraints on $\Lambda$ to scale approximately with the square root of the integrated luminosity. At future colliders, such as FCC, a larger centre-of-mass energy implies that more resonances become kinematically accessible, so the impact of the tower is likely to be even more significant than what is considered here.

\section*{Acknowledgments}
The authors thank I.~Martinez~Soler and D.~Pasari for useful discussions. We would also like to thank Jie Xiao, Mattia Campana, Jan-Frederik Schulte, Hsin-Yeh Wu and Antonis Agapitos for correspondence about mass-resolution models used in the CMS analyses. MM is supported by a Royal Society Newton International Fellowship (NIF/R1/232539). FS is supported by the STFC under grant agreement ST/P006744/1.
\appendix
\section{Constraints in the Randall-Sundrum Model}
\label{app:RS-model}

In the Randall-Sundrum model~\cite{Randall:1999vf,Randall:1999ee}, a slice of AdS$_5$ is compactified on the orbifold $S^1/Z_2$, with metric
\begin{equation}
    ds^2 = e^{-2 k r_c |\phi|}\,\eta_{\mu\nu}dx^\mu dx^\nu - r_c^2\, d\phi^2\,, \qquad \phi\in[-\pi,\pi]\,,
    \label{eq:RS-metric}
\end{equation}
where $k$ is the AdS$_5$ curvature scale and $r_c$ the radius of the extra dimension, with two branes sitting at the orbifold fixed points $\phi=0,\pi$; the warping is controlled by the dimensionless combination $k r_c$. Indeed, in the two cases, the masses of the 4D-EFT spectrum read
\begin{align}
    m_n &= \frac{n}{r_c}\left(1-\frac{\pi}{2}\,k r_c\right)+\mathcal{O}\!\left((k r_c)^2\right)\,, &  k r_c &\ll 1\,, \label{eq:mn-smallk}\\
    m_n &= \gamma_n k\, e^{-k r_c \pi}\approx \pi\left(n+\frac14\right) k\, e^{-k r_c \pi}+\mathcal{O}\!\left(n^{-1}\right)\,, &  k r_c &\gg 1\,, \label{eq:mn-largek}
\end{align}
where \eqref{eq:mn-largek} follows from the large-argument behaviour of the roots of the Bessel function $J_1$, $J_1(\gamma_n)=0$, which control the exact RS graviton spectrum~\cite{Davoudiasl:1999jd}.

In the small-curvature limit, the $n$-dependence factorises exactly at $\mathcal{O}(k r_c)$: every KK level is rescaled by the same $n$-independent factor, and thus the mass ratios are left completely unchanged with respect to the flat case. Consequently, at this order in $kr_c$, our benchmark spectrum $m_n=n\,m_1$ is reproduced \emph{exactly} by the identification
\begin{equation}
    m_1 \equiv \frac{1}{r_c}\left(1-\frac{\pi}{2}\,k r_c\right)\,,
    \label{eq:m1-smallk}
\end{equation}
and the constraints derived in the main text can be interpreted accordingly.

\begin{figure}
    \centering
    \includegraphics[width=\linewidth]{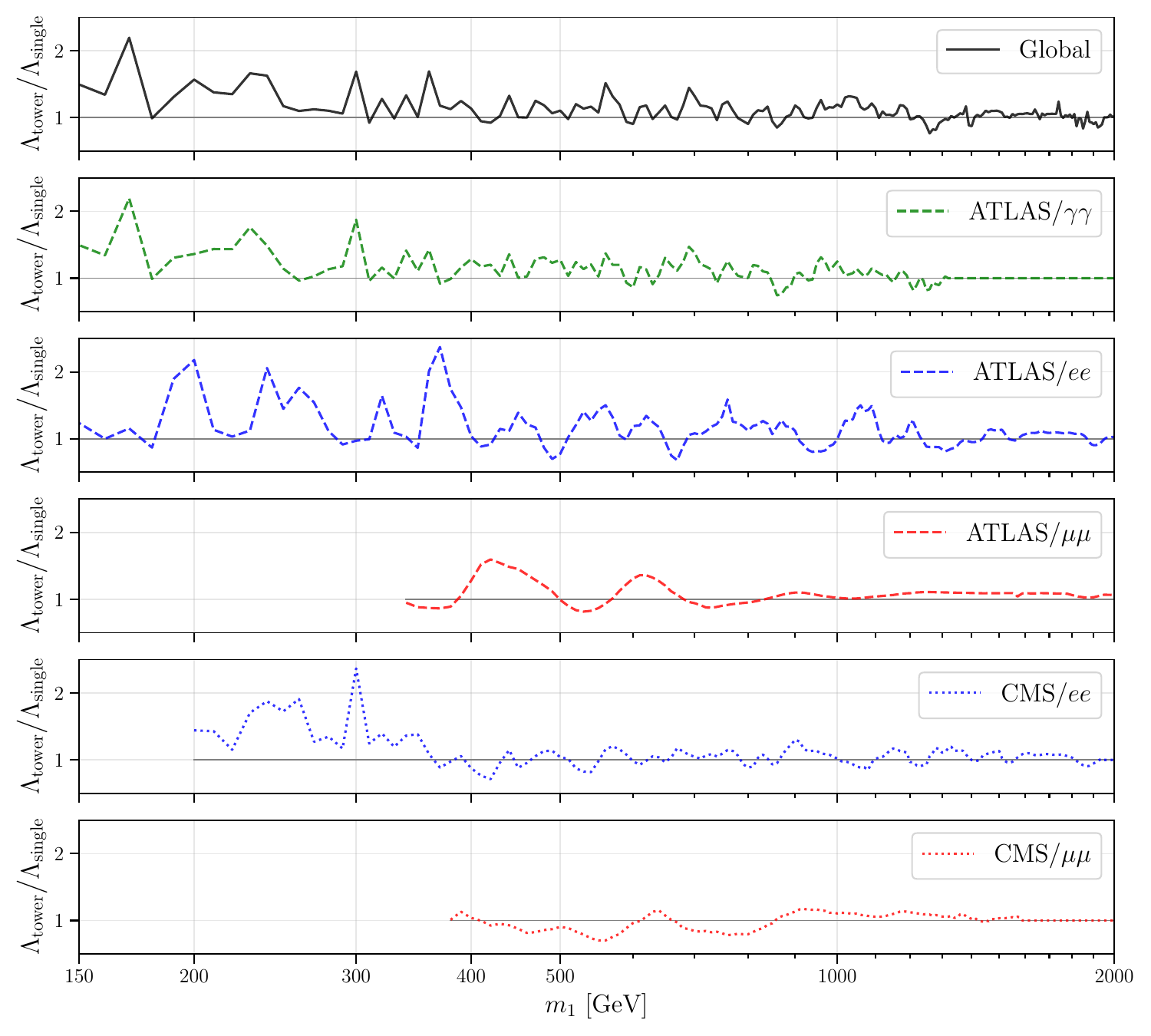}
        \caption{Ratio of the bounds at the $95\%$~C.L. derived including the whole tower of RS gravitons ($\Lambda_\text{tower}$) with respect to the one with the first RS graviton ($\Lambda_\text{single}$) for different datasets of ATLAS and CMS.
        }
        \label{fig:exact-spectrum-ratio}
\end{figure}

In the large-curvature limit, it is the level splitting itself that becomes $n$-independent, $m_{n+1}-m_n\approx\pi k\, e^{-kr_c\pi}$, the defining property of $m_n=n\,m_1$. Identifying $m_1$ with this splitting, the spectrum is
\begin{equation}
    m_n = m_1\left(n+\frac14\right)+\mathcal{O}(n^{-1})\,,
\end{equation}
with the $\mathcal{O}(n^{-1})$ correction to this asymptotic formula already negligible from $n=2$. The approximation fails at the level of $20\%$ for the first mode. 
\begin{figure}
    \centering
    \includegraphics[width=\linewidth]{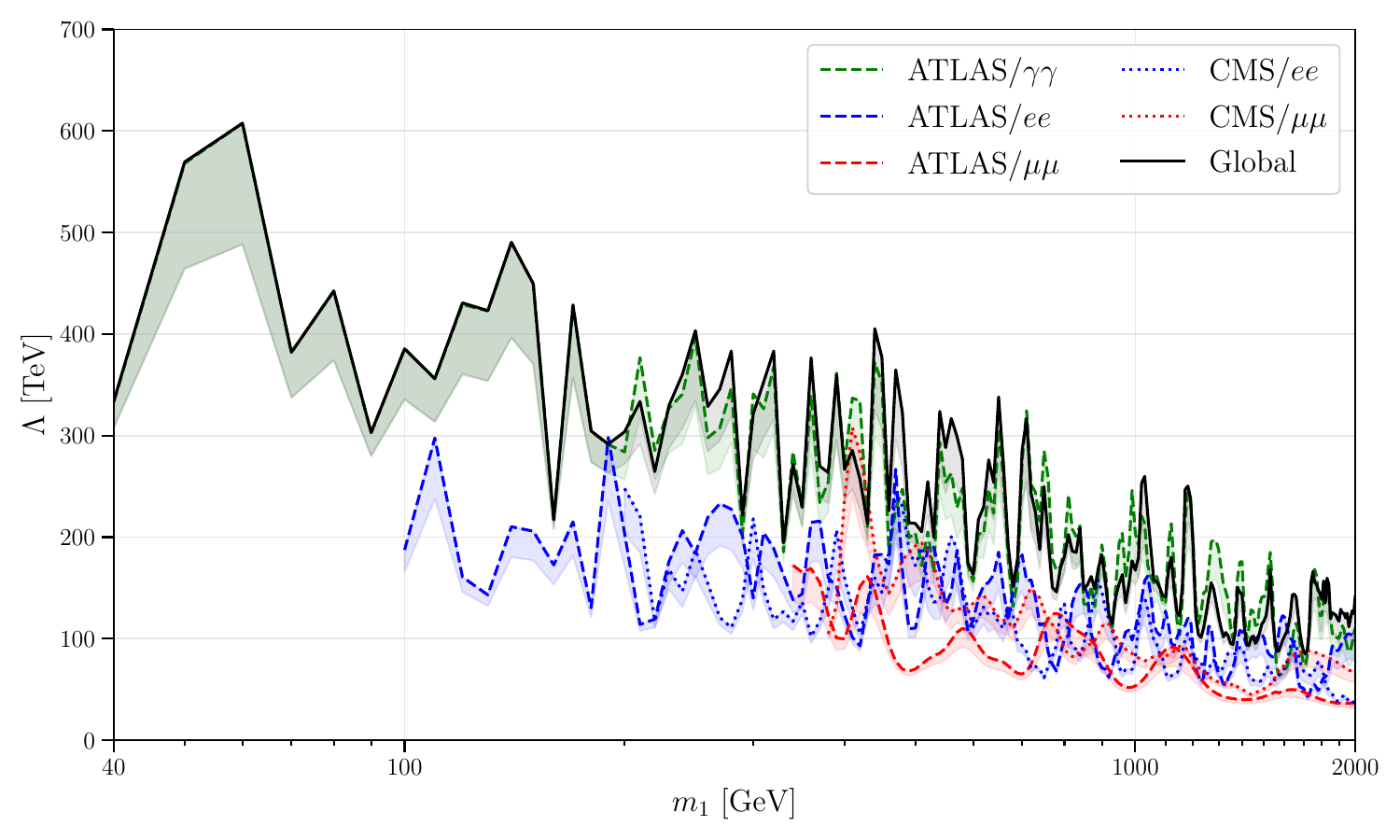}
    \caption{Lower bound at the $95\%$~C.L. on the interaction scale $\Lambda$ including the whole tower of RS gravitons employing data from ATLAS~(dashed) and CMS~(dotted) including the diphoton~(green), dielectron~(blue) and dimuon~(red) final states. The comprehensive constraint derived from the combination of the different dataset likelihoods is shown with a solid black line. The lower edge of the shaded band indicates bounds at the $99\%$~C.L.~.}
    \label{fig:exact-spectrum-maximal}
\end{figure}
The constraints for this case are shown in Fig.~\ref{fig:exact-spectrum-maximal}. Since the modes that contribute the most are the light ones, and since the bins are dense compared to the distance between two massive modes, the resulting constraints do feature differences with respect to the phenomenological benchmark model considered in Section~\ref{sec:models}. This is particularly visible when comparing the relative enhancement of the tower compared to the lightest mode of Fig.~\ref{fig:exact-spectrum-ratio} with Fig.~\ref{fig:KK-constraints-ratio} from the main text.

\section{KK Graviton Decay Channels}
In this Appendix, we report the decay widths of a graviton into SM states and into other gravitons within the RS model. The decay channels of massive gravitons into SM have been computed in the literature, e.g., in Ref.~\cite{deGiorgi:2021xvm}.

\subsection{Graviton to Standard Model}
\label{app:width-SM}
\begin{figure*}
    \centering
    \includegraphics[width=\linewidth]{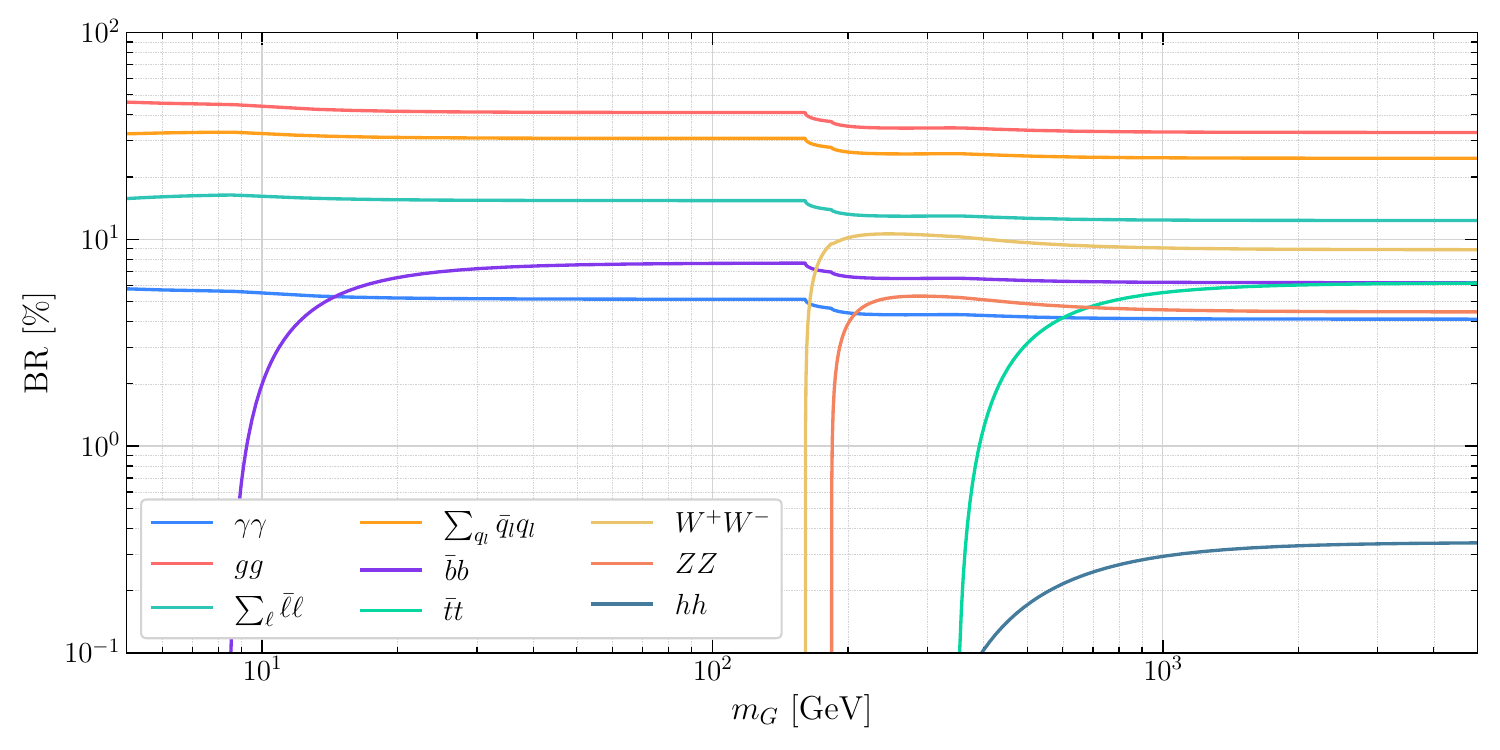}
    \caption{Branching ratios of a massive graviton into SM final states. The sum over $q_l$ includes only quarks lighter than the $b$-quark, while the sum over leptons $\ell$ includes both charged ones and neutrinos.}
    \label{fig:single-branching}
\end{figure*}

We label $m_n$ the mass of a graviton $G_n$ and define for convenience the dimensionless ratio
\begin{equation}
    x_y\equiv \left(\frac{2m_y}{m_n}\right)^2\,.
\end{equation}
The decay widths of $G_n$ into SM states read:
\begin{align}
    &\Gamma\left(hh\right)= \frac{m_n^3}{960 \pi  \Lambda ^2} \left(1-x_h\right)^{5/2} \,,\\
    &\Gamma\left(\Bar{\psi}\psi \right)= \frac{m_n^3}{160 \pi  \Lambda ^2 }\left(1-x_\psi\right)^{\frac{3}{2}} \left(1+\frac{2}{3}x_\psi\right)\,,\\
    &\Gamma\left(Z Z\right)=\frac{m_n^3}{960 \pi  \Lambda ^2}\left(1-x_Z\right)^{\frac{1}{2}} \left(13+14x_Z+3 x_Z^2\right)\,,\\
    &\Gamma\left( W_+\  W_-\right)=\frac{m_n^3}{480 \pi  \Lambda ^2}\left(1-x_W\right)^{\frac{1}{2}} \left(13+14x_W+3 x_W^2\right)\,,\\
    & \Gamma\left(\gamma  \gamma\right)=\frac{m_n^3}{80 \pi  \Lambda ^2}\,,\\
    & \Gamma\left(gg\right)=\frac{m_n^3}{10 \pi  \Lambda ^2}\,.
\end{align}
where $\bar{\psi}\psi$ is a generic SM singlet fermion-antifermion pair. 
The branching ratios as a function of the graviton mass can be seen in Fig.~\ref{fig:single-branching}.
Summing over all decay channels with appropriate colour factors, and assuming $m_n\gg 2m_t$, with $m_t$ being the mass of the top quark, one finds
\begin{equation}
    \Gamma(G_n\to\text{SM})\approx\frac{583}{1920\pi}\frac{m_n^3}{\Lambda^2} \approx  0.097\frac{m_n^3}{\Lambda^2} \ .
\end{equation}

\subsection{Graviton to Gravitons}
\label{app:width-gravitons}
The structure of the widths for graviton final states is rather complicated if compared to the SM ones.
The model dependence of the interaction strength between three gravitons $G_n G_m G_k$ is encoded into dimensionless coefficients $\chi_{nmk}$.
Within the RS model, the value of the coefficients $\chi_{nmk}$ is given by~\cite{deGiorgi:2021xvm}
\begin{equation}
    \chi_{ijk}\equiv \frac{-2}{J_0(\gamma_i)J_0(\gamma_j)J_0(\gamma_k)}\int\limits_0^1 \diff{u} \ u^3 J_2(\gamma_iu)J_2(\gamma_ju)J_2(\gamma_ku)\,,
\end{equation}
where $J_n$ are J-Bessel functions and $\gamma_n$ is the nth-zero of the $J_1$ Bessel function. By comparison, in the flat case $\chi_{ijk}=\sqrt{2}$ if $i\pm j\pm k=0$, and zero otherwise.

The general expression for the decay into two lighter gravitons $G_m$ and $G_k$ is
\begin{align}
     & \nonumber\Gamma_{G_n\rightarrow G_m G_k}=\frac{\chi _{\text{nkm}}^2}{17280 \pi  \Lambda^2}  \left[
\left(m_n^2-\left(m_k-m_m\right)^2\right)\left(m_n^2-\left(m_k+m_m\right)^2\right)\right]{}^{5/2}\times  \\
     & \times  \left[14 m_k^4 \left(26 m_m^2 m_n^2+9 m_m^4+9 m_n^4\right)+ 26 m_k^2 \left(14 m_m^4 m_n^2+14 m_m^2 m_n^4+m_m^6+m_n^6\right)+\right. \\
     &\nonumber \left. +m_k^8+26 m_m^2 m_n^6+126 m_m^4 m_n^4 +26 m_m^6 m_n^2+m_m^8+m_n^8+26 m_k^6 \left(m_m^2+m_n^2\right)\right]/(m_k^4 m_m^4 m_n^7) \,,\\
    & \Gamma_{G_n  \rightarrow  G_m G_m}=\frac{\chi _{\text{nmm}}^2 \left(m_n^2-4 m_m^2\right){}^{5/2} \left(780 m_m^6 m_n^2+616 m_m^4 m_n^4+52 m_m^2 m_n^6+180 m_m^8+m_n^8\right)}{34560 \pi  \Lambda ^2 m_m^8 m_n^2}\,.
\end{align}

\subsection{Comparison between SM and Gravitons Final States}
\label{app:comparison-widths}

In this Appendix, we compare the branching ratios of a massive graviton into two massive KK gravitons to SM final states. We employ the formulas of Apps.~\ref{app:width-SM}-\ref{app:width-gravitons}.

The decay width of gravitons into any pair of graviton is given by
\begin{align}
    \label{eq:KK-width}\Gamma(G_n\to GG)\equiv\sum\limits_{j\leq i}\Gamma(G_n\to G_iG_j)\,, 
\end{align}
\begin{figure}
    \centering
    \includegraphics[width=\linewidth]{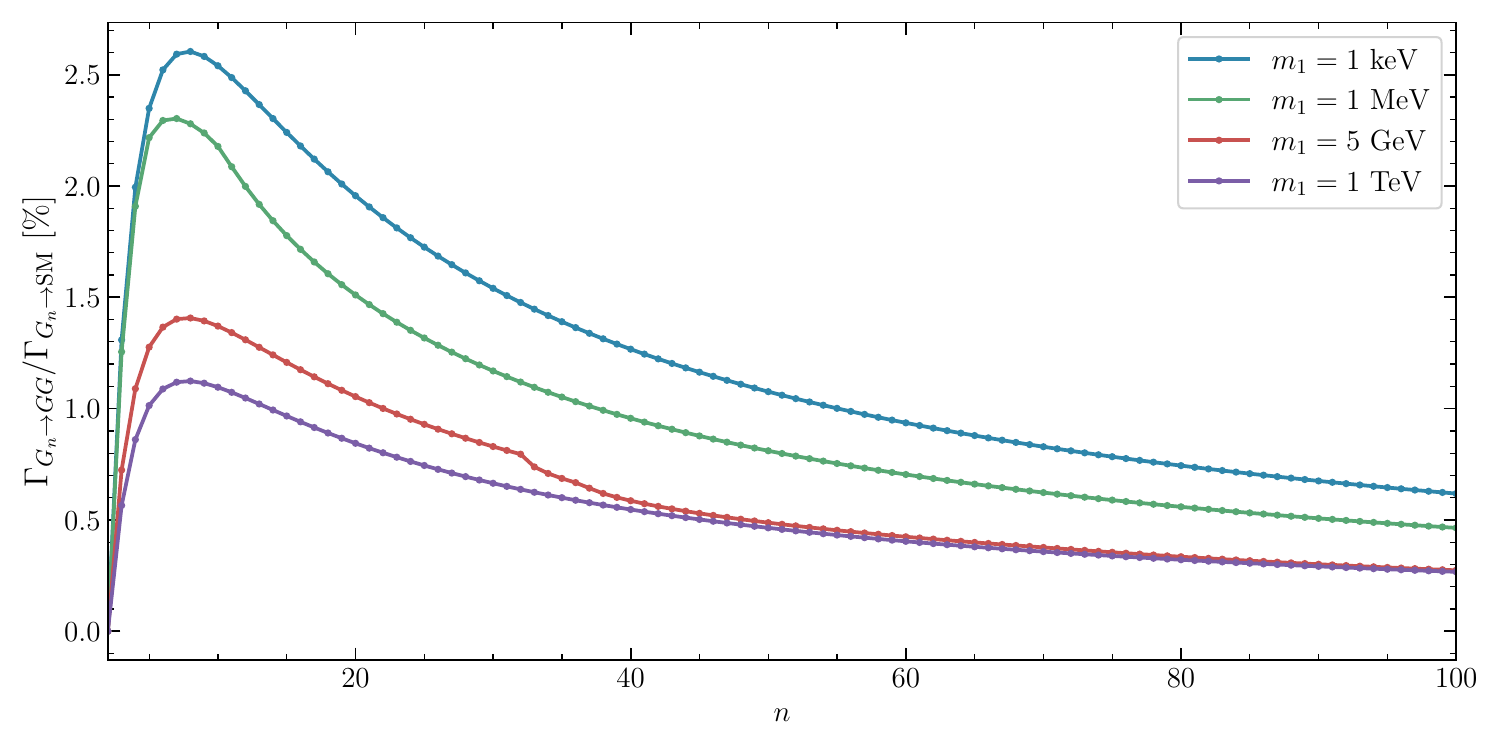}
    \caption{Ratio of the decay width of a graviton to lighter KK modes over SM decay channels, $\Gamma(G_n\to GG)/\Gamma(G_n\to \text{SM})$, as defined in Eq.~\eqref{eq:KK-width} as a function of the decaying graviton mass $m_n=nm_1$ for benchmark values of $m_1$. }
    \label{fig:KK-branching}
\end{figure}
In the limit where $m_n \gg 2m_t$, the ratio of the graviton width to SM final states, $\Gamma_{G_n\to\text{SM}~\text{SM}}$ becomes independent of $m_1$ as the overall mass scale cancels out.
Its exact functional dependence can be seen in Fig.~\ref{fig:KK-branching}.  For completeness, we report both the case of heavy gravitons (useful to this work) and modes as light as a keV. The change of slope for the $m_1=5$~GeV case between $n\sim 30-40$ is due to the increase in the SM width due to the opening of the $W,Z,h, t$ channels.
As can be seen, for a TeV~graviton, the ratio is always smaller than $0.01$, it peaks around $n=8$, and then it monotonically decreases. 
As $m_1$ decreases, the SM gets smaller due to the closure of decay channels. For $m_1$ as light as a keV, the ratio reaches at most a $2.5\%$. For all masses, its value decreases as $n$ increases.
The cause for such a somewhat counter-intuitive dependence lies in the size of the phase space, jointly with the structure of the couplings.
All in all, this allows us to neglect graviton final states in our computations. 

\section{Systematics and Nuisance parameters}
\label{app:systematics}

In this appendix, we summarise the systematic effects that we encode in the nuisance parameters of our statistical method, described in Sec.~\ref{sec:method}. 
\begin{table}[h]
\centering
\small
\setlength{\tabcolsep}{4pt}
\begin{tabular}{|p{0.2\textwidth}p{0.8\textwidth}|}
\hline
Dataset  &  $\delta^k(\mathrm{m_G})\, [\%] $  \\
\hline
\hline
ATLAS $ee$ & Spurious signal: $12.5 (300 \,\mathrm{GeV} )\,,\,4.6 (2000 \,\mathrm{GeV}) \,,\,1.0 (3000 \,\mathrm{GeV})$\\
         &  Luminosity: $1.7$\\
         &  Lepton id: $1.6 (300 \,\mathrm{GeV}) \,,\, 5.6 (2000 \,\mathrm{GeV}) \,,\, 5.6 (5000 \,\mathrm{GeV})$\\
         &  Isolation: $0.3 (300 \,\mathrm{GeV}) \,,\, 1.1 (2000 \,\mathrm{GeV}) \,,\, 1.1 (5000 \,\mathrm{GeV})$\\
ATLAS $\mu\mu$ & Spurious signal: $11.7 (300 \,\mathrm{GeV})\,,\,3.5(2000 \,\mathrm{GeV})\,,\,2.1 (5000 \,\mathrm{GeV})$\\
               & Lepton id: $1.8 (300 \,\mathrm{GeV}) \,,\, 11 (2000 \,\mathrm{GeV}) \,,\,22.5 (5000 \,\mathrm{GeV})$\\
               & Isolation: $4$ \\
               & Luminosity: $1.7$\\
               & Good Muons: $0.6 (300 \,\mathrm{GeV}) \,,\,8.6 (2000 \,\mathrm{GeV})\,,\,45 (5000 \,\mathrm{GeV})$. \\
ATLAS $\gamma\gamma$ & Luminosity: $1.7$\\
                     & Photon identification: $0.5$\\
                     & Trigger: $0.5$\\
                     & Photon isolation: $1.5$\\
                     & Pileup reweighting: $2.0 (160 \,\mathrm{GeV}) \,,\, 0.2 (2800 \,\mathrm{GeV})$\\
CMS $ee$ & Electron selection efficiency only: $7$\\
CMS $\mu\mu$ & Muon selection eff.\ (double): $2$\\
             & Muon selection eff.\ (single): $2$\\
\hline
\end{tabular}
\normalsize
\caption{Summary of systematics included as nuisance parameters in the likelihood profiling.}
\label{tab:nuisance-config-summary}
\end{table}
Table~\ref{tab:nuisance-config-summary} contains their relative sizes $\delta^k$, which are applied to the graviton signal yield, with one independent nuisance parameter each. The label $k$ indicates the possible sources of systematic uncertainties.
We quote them directly from the corresponding experimental analysis papers~\cite{ATLAS:2019erb,CMS:2021ctt,ATLAS:2021uiz}.

For a specific source of systematics, when the size $\delta^k$ is given at individual benchmark mass points, we interpolate the shift $\delta^k(m_G)$ between known values and apply it as a correction to the total graviton yield.
\begin{table}[h]
\centering
\small
\setlength{\tabcolsep}{4pt}
\begin{tabular}{|p{0.2\textwidth}p{0.8\textwidth}|}
\hline
Dataset  & $r^{\mathrm{res}} (m_G)\, [\%]$  \\
\hline
\hline
ATLAS $\gamma\gamma$ & $10 (160 \,\mathrm{GeV}) \,,\, 50  (2800 \,\mathrm{GeV})$\\
ATLAS $\mu\mu$ & $3.8 (300 \,\mathrm{GeV}) \,,\, 3.2 (2000 \,\mathrm{GeV}) \,,\, 2.4 (5000 \,\mathrm{GeV})$ \\
ATLAS ee &  $8.3 (300 \,\mathrm{GeV}) \,,\, 11.8  (2000 \,\mathrm{GeV}) \,,\, 9.0 (5000 \,\mathrm{GeV})$\\
CMS $\mu\mu$ & $15$ \\
\hline
\end{tabular}
\normalsize
\caption{Energy-resolution systematic. The variation acts multiplicatively on the mass-resolution model spread.}
\label{tab:resolution-systematics}
\end{table}
The detector energy-resolution systematic is handled separately by template morphing. In practice, for each mass bin $i$, we consider the signal $s_i$ generated with the nominal value of the mass-resolution model parameters, and a modified signal $s_i^{\rm res}$ obtained by increasing the spread of the mass-resolution model by a factor $r^{\rm res}$, as given in Table~\ref{tab:resolution-systematics}. The two templates are then combined to build a bin-wise fractional shift
\begin{equation}
    \forall s_i \neq 0, \quad \delta_i^{\mathrm{res}}=\frac{s_i^{\mathrm{res}}-s_i}{s_i},
\end{equation}
which is then applied with a linear nuisance parameter on top of the nominal signal.

\section{Comparison with the Results of the Collaborations}
\label{app:comparison}

This appendix provides a detailed comparison between the constraints derived from our analysis and the ones reported or extracted from ATLAS and CMS.
\subsection{Comparison with ATLAS}
\label{app:comparison-ATLAS}
We compare here the recast from the ATLAS dilepton~\cite{ATLAS:2019erb} and diphoton~\cite{ATLAS:2021uiz} searches with our results. This step is essential to calibrate the reliability and limits of our results. 

The ATLAS collaboration provides the 95\%~C.L. on the total production and decay cross-section $\sigma(pp\to G\to ee/\mu\mu/\gamma\gamma)$, which we translate to bounds in the $(m_G,\Lambda)$ parameter space. However, in~\cite{ATLAS:2021uiz} the upper bound is presented solely for values of $k/\overline{M}_\text{P} \geq 0.01$. To extend the exclusion region below this threshold, we adopt the last available upper bound at 
$k/\overline{M}_\text{P} = 0.01$ and apply it uniformly for all $k/\overline{M}_\text{P} < 0.01$. While this treatment is not strictly rigorous, it is expected to be a reasonable approximation: in this regime, the resonance width becomes increasingly small, eventually falling below the experimental energy resolution. When this happens, smearing effects of the detector start to dominate, thus washing out signal differences for different ratios of $k/\overline{M}_\text{P}$.

\begin{figure}
    \centering
    \includegraphics[width=\linewidth]{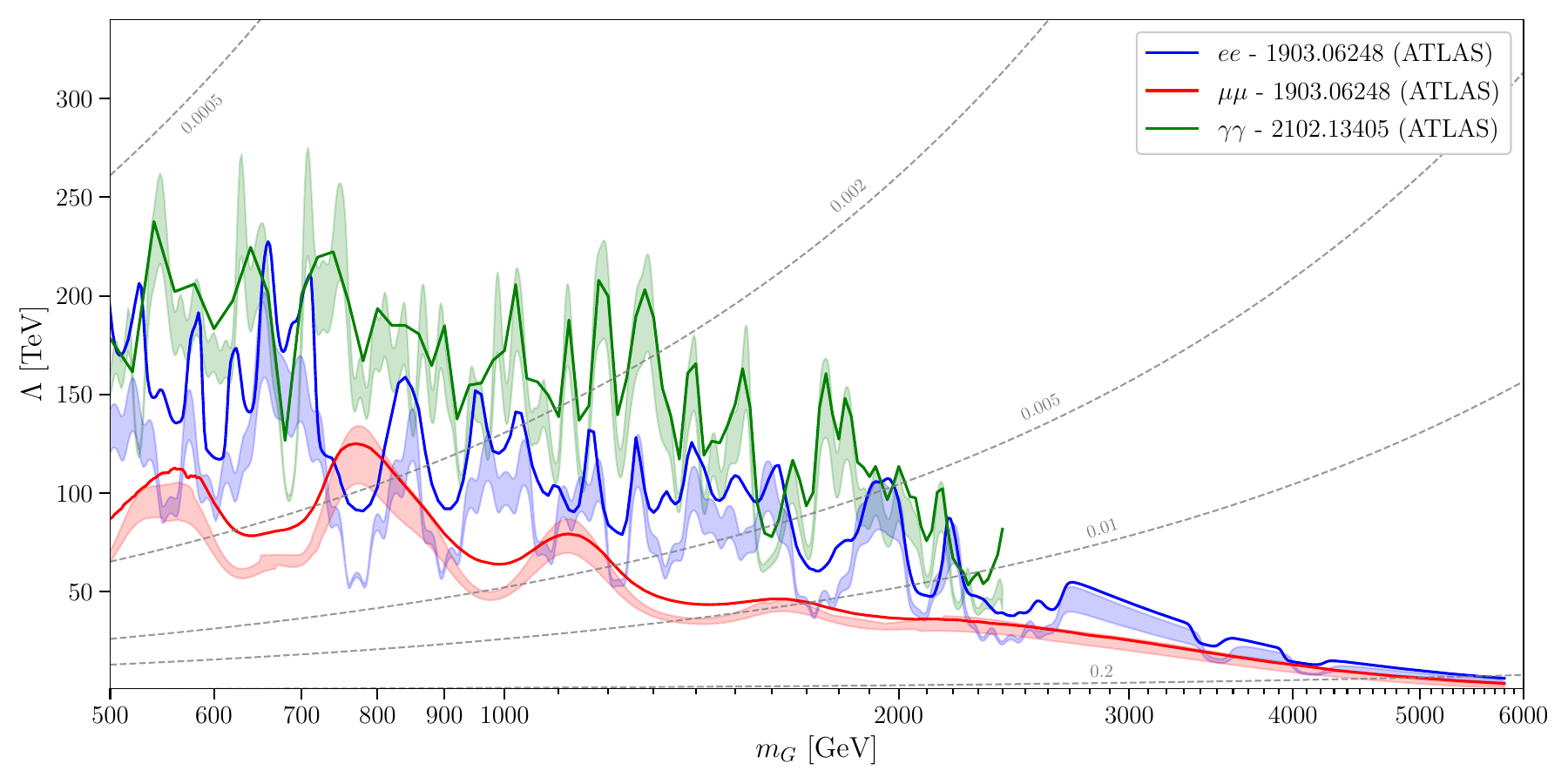}
    \caption{Comparison between the constraints derived in this work at $95\%$ and $99\%$~C.L. for the single resonance scenario with the ones derived by ATLAS at $95\%$~C.L. Contours of different values of $k/\overline{M}_P$ are shown with dashed gray lines.}
    \label{fig:comparison-ATLAS}
\end{figure}
The comparison between our results and the bounds we directly extracted from Ref.~\cite{ATLAS:2019erb,ATLAS:2021uiz} is shown in Fig.~\ref{fig:comparison-ATLAS}.
The figure shows our bounds derived at 95\%  and 99\%~C.L. (upper and lower lines of the shaded area) and the limits directly extracted from the same references at 95\%~C.L.\ (solid lines). 



\subsection{Comparison with CMS}
\label{app:comparison-CMS}

We compare here the recast from the CMS dilepton~\cite{CMS:2021ctt} searches with our results. This is necessary to validate our results.
\begin{figure}
    \centering
    \includegraphics[width=\linewidth]{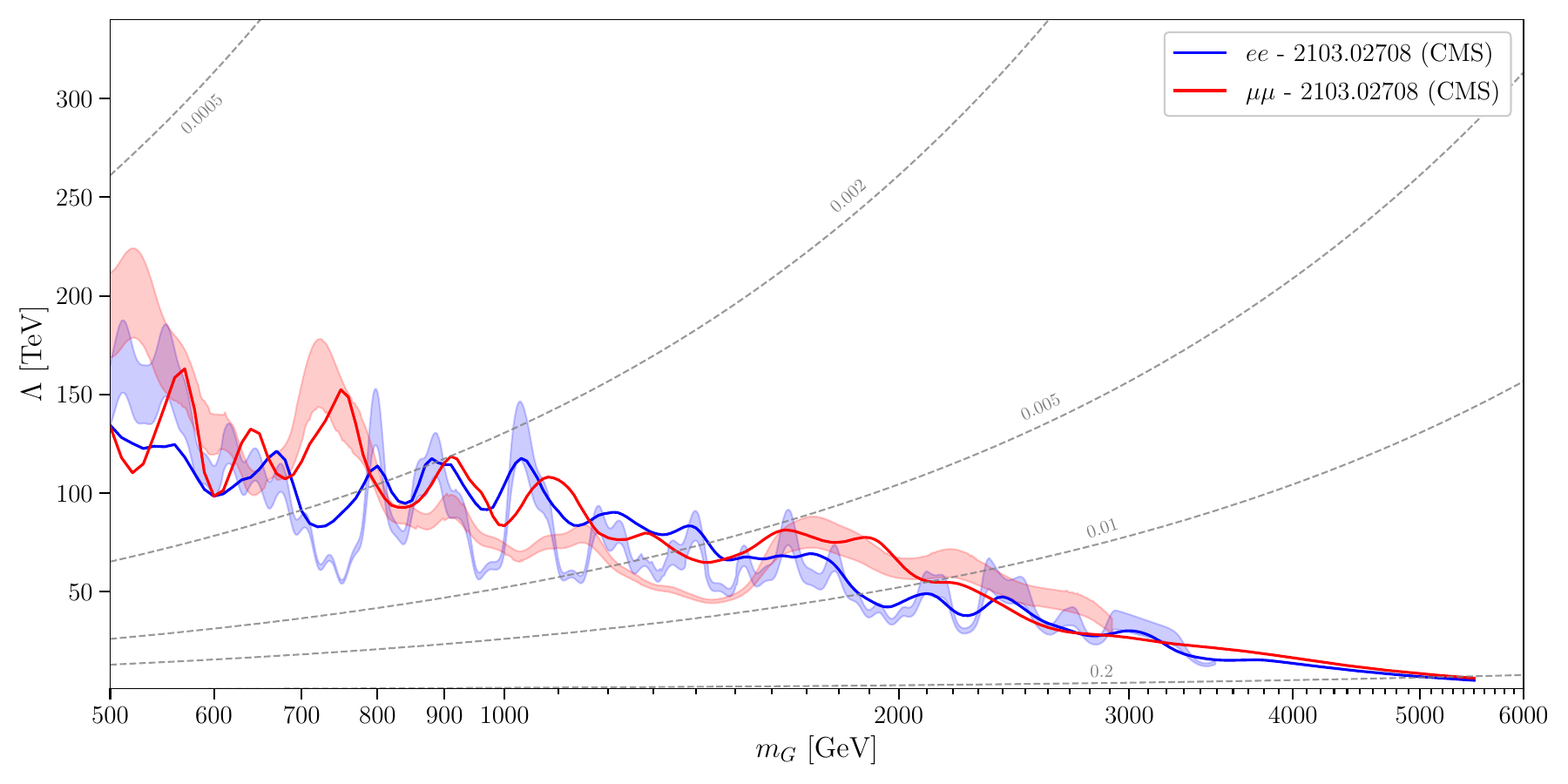}
    \caption{Comparison between the constraints derived in this work at $95\%$ and $99\%$~C.L. for the single resonance scenario with the ones derived by CMS at $95\%$~C.L. Contours of different values of $k/\overline{M}_P$ are shown with dashed gray lines.}
    \label{fig:comparison-CMS}
\end{figure}

The comparison is shown in Fig.~\ref{fig:comparison-CMS}, where the shaded bands display our bounds at $95\%$ and $99\%$~C.L. and the solid lines are the limits directly extracted from Ref.~\cite{CMS:2021ctt} at $95\%$~C.L., for both the $ee$ and $\mu\mu$ final states.
The overall envelope of our constraints agrees well with the published CMS limits across the full mass range, for both channels. Local fluctuations are present, with the two curves alternately crossing each other, but no systematic bias in either direction is observed. 
\bibliographystyle{BiblioStyle}
\bibliography{references}

@article{Kaluza:1921tu,
    author = "Kaluza, Th.",
    title = {{Zum Unit{\"a}tsproblem der Physik}},
    eprint = "1803.08616",
    archivePrefix = "arXiv",
    primaryClass = "physics.hist-ph",
    reportNumber = "HUPD-8401",
    doi = "10.1142/S0218271818700017",
    journal = "Sitzungsber. Preuss. Akad. Wiss. Berlin (Math. Phys. )",
    volume = "1921",
    pages = "966--972",
    year = "1921"
}

@article{Klein:1926tv,
    author = "Klein, Oskar",
    editor = "Taylor, J. C.",
    title = "{Quantum Theory and Five-Dimensional Theory of Relativity. (In German and English)}",
    doi = "10.1007/BF01397481",
    journal = "Z. Phys.",
    volume = "37",
    pages = "895--906",
    year = "1926"
}

@article{Davoudiasl:1999jd,
    author = "Davoudiasl, H. and Hewett, J. L. and Rizzo, T. G.",
    title = "{Phenomenology of the Randall-Sundrum Gauge Hierarchy Model}",
    eprint = "hep-ph/9909255",
    archivePrefix = "arXiv",
    reportNumber = "SLAC-PUB-8241",
    doi = "10.1103/PhysRevLett.84.2080",
    journal = "Phys. Rev. Lett.",
    volume = "84",
    pages = "2080",
    year = "2000"
}

@article{Antoniadis:1990ew,
    author = "Antoniadis, Ignatios",
    title = "{A Possible new dimension at a few TeV}",
    reportNumber = "EP-CPTH-A978-0690",
    doi = "10.1016/0370-2693(90)90617-F",
    journal = "Phys. Lett. B",
    volume = "246",
    pages = "377--384",
    year = "1990"
}

@article{Arkani-Hamed:1998jmv,
    author = "Arkani-Hamed, Nima and Dimopoulos, Savas and Dvali, G. R.",
    title = "{The Hierarchy problem and new dimensions at a millimeter}",
    eprint = "hep-ph/9803315",
    archivePrefix = "arXiv",
    reportNumber = "SLAC-PUB-7769, SU-ITP-98-13",
    doi = "10.1016/S0370-2693(98)00466-3",
    journal = "Phys. Lett. B",
    volume = "429",
    pages = "263--272",
    year = "1998"
}

@article{Randall:1999ee,
    author = "Randall, Lisa and Sundrum, Raman",
    title = "{A Large mass hierarchy from a small extra dimension}",
    eprint = "hep-ph/9905221",
    archivePrefix = "arXiv",
    reportNumber = "MIT-CTP-2860, PUPT-1860, BUHEP-99-9",
    doi = "10.1103/PhysRevLett.83.3370",
    journal = "Phys. Rev. Lett.",
    volume = "83",
    pages = "3370--3373",
    year = "1999"
}

@article{Randall:1999vf,
    author = "Randall, Lisa and Sundrum, Raman",
    title = "{An Alternative to compactification}",
    eprint = "hep-th/9906064",
    archivePrefix = "arXiv",
    reportNumber = "MIT-CTP-2874, PUPT-1867, BUHEP-99-13",
    doi = "10.1103/PhysRevLett.83.4690",
    journal = "Phys. Rev. Lett.",
    volume = "83",
    pages = "4690--4693",
    year = "1999"
}

@article{Giudice:2016yja,
    author = "Giudice, Gian F. and McCullough, Matthew",
    title = "{A Clockwork Theory}",
    eprint = "1610.07962",
    archivePrefix = "arXiv",
    primaryClass = "hep-ph",
    reportNumber = "CERN-TH-2016-223",
    doi = "10.1007/JHEP02(2017)036",
    journal = "JHEP",
    volume = "02",
    pages = "036",
    year = "2017"
}

@article{deGiorgi:2021xvm,
    author = "de Giorgi, Arturo and Vogl, Stefan",
    title = "{Dark matter interacting via a massive spin-2 mediator in warped extra-dimensions}",
    eprint = "2105.06794",
    archivePrefix = "arXiv",
    primaryClass = "hep-ph",
    doi = "10.1007/JHEP11(2021)036",
    journal = "JHEP",
    volume = "11",
    pages = "036",
    year = "2021"
}

@article{ATLAS:2023hbp,
    author = "Aad, Georges and others",
    collaboration = "ATLAS",
    title = "{Search for periodic signals in the dielectron and diphoton invariant mass spectra using 139 fb$^{-1}$ of pp collisions at $ \sqrt{s} $ = 13 TeV with the ATLAS detector}",
    eprint = "2305.10894",
    archivePrefix = "arXiv",
    primaryClass = "hep-ex",
    reportNumber = "CERN-EP-2023-073",
    doi = "10.1007/JHEP10(2023)079",
    journal = "JHEP",
    volume = "10",
    pages = "079",
    year = "2023"
}

@article{ATLAS:2019erb,
    author = "Aad, Georges and others",
    collaboration = "ATLAS",
    title = "{Search for high-mass dilepton resonances using 139 fb$^{-1}$ of $pp$ collision data collected at $\sqrt{s}=$13 TeV with the ATLAS detector}",
    eprint = "1903.06248",
    archivePrefix = "arXiv",
    primaryClass = "hep-ex",
    reportNumber = "CERN-EP-2019-030",
    doi = "10.1016/j.physletb.2019.07.016",
    journal = "Phys. Lett. B",
    volume = "796",
    pages = "68--87",
    year = "2019"
}

@article{ATLAS:2021uiz,
    author = "Aad, Georges and others",
    collaboration = "ATLAS",
    title = "{Search for resonances decaying into photon pairs in 139 fb$^{-1}$ of $pp$ collisions at $\sqrt {s}$=13 TeV with the ATLAS detector}",
    eprint = "2102.13405",
    archivePrefix = "arXiv",
    primaryClass = "hep-ex",
    reportNumber = "CERN-EP-2020-248",
    doi = "10.1016/j.physletb.2021.136651",
    journal = "Phys. Lett. B",
    volume = "822",
    pages = "136651",
    year = "2021"
}

@article{CMS:2021ctt,
    author = "Sirunyan, Albert M and others",
    collaboration = "CMS",
    title = "{Search for resonant and nonresonant new phenomena in high-mass dilepton final states at $ \sqrt{s} $ = 13 TeV}",
    eprint = "2103.02708",
    archivePrefix = "arXiv",
    primaryClass = "hep-ex",
    reportNumber = "CMS-EXO-19-019, CERN-EP-2021-026",
    doi = "10.1007/JHEP07(2021)208",
    journal = "JHEP",
    volume = "07",
    pages = "208",
    year = "2021"
}

@article{CMS:2024nht,
    author = "Hayrapetyan, Aram and others",
    collaboration = "CMS",
    title = "{Search for new physics in high-mass diphoton events from proton-proton collisions at $ \sqrt{\textrm{s}} $ = 13 TeV}",
    eprint = "2405.09320",
    archivePrefix = "arXiv",
    primaryClass = "hep-ex",
    reportNumber = "CMS-EXO-22-024, CERN-EP-2024-109",
    doi = "10.1007/JHEP08(2024)215",
    journal = "JHEP",
    volume = "08",
    pages = "215",
    year = "2024"
}

@article{Junk:1999kv,
    author = "Junk, Thomas",
    title = "{Confidence level computation for combining searches with small statistics}",
    eprint = "hep-ex/9902006",
    archivePrefix = "arXiv",
    reportNumber = "CARLETON-OPAL-PHYS-99-01, CERN-EP-99-041",
    doi = "10.1016/S0168-9002(99)00498-2",
    journal = "Nucl. Instrum. Meth. A",
    volume = "434",
    pages = "435--443",
    year = "1999"
}

@article{Wilks:1938dza,
    author = "Wilks, S. S.",
    title = "{The Large-Sample Distribution of the Likelihood Ratio for Testing Composite Hypotheses}",
    doi = "10.1214/aoms/1177732360",
    journal = "Annals Math. Statist.",
    volume = "9",
    number = "1",
    pages = "60--62",
    year = "1938"
}

@article{deGiorgi:2022yha,
    author = "de Giorgi, Arturo and Vogl, Stefan",
    title = "{Warm dark matter from a gravitational freeze-in in extra dimensions}",
    eprint = "2208.03153",
    archivePrefix = "arXiv",
    primaryClass = "hep-ph",
    doi = "10.1007/JHEP04(2023)032",
    journal = "JHEP",
    volume = "04",
    pages = "032",
    year = "2023"
}

@article{deGiorgi:2025xgp,
    author = "de Giorgi, Arturo and Pasari, Dhruv and Turner, Jessica",
    title = "{Do neutrinos dream in 5D? Towards a comprehensive extra-dimensional neutrino phenomenology}",
    eprint = "2512.02101",
    archivePrefix = "arXiv",
    primaryClass = "hep-ph",
    reportNumber = "IPPP/25/84",
    doi = "10.1007/JHEP05(2026)152",
    journal = "JHEP",
    volume = "05",
    pages = "152",
    year = "2026"
}

@article{Donini:2025qrf,
    author = "Donini, Andrea and Folgado, Miguel G. and Mu{\~n}oz-Ovalle, Alejandro",
    title = "{Dark matter in a three-brane Randall-Sundrum scenario out of the evanescent limit}",
    eprint = "2509.04580",
    archivePrefix = "arXiv",
    primaryClass = "hep-ph",
    doi = "10.1007/JHEP02(2026)206",
    journal = "JHEP",
    volume = "02",
    pages = "206",
    year = "2026"
}

@article{Donini:2025cpl,
    author = "Donini, Andrea and Folgado, Miguel G. and Herrero-Garc{\'\i}a, Juan and Landini, Giacomo and Mu{\~n}oz-Ovalle, Alejandro and Rius, Nuria",
    title = "{Dark Matter in an evanescent three-brane Randall-Sundrum scenario}",
    eprint = "2505.13601",
    archivePrefix = "arXiv",
    primaryClass = "hep-ph",
    doi = "10.1007/JHEP11(2025)037",
    journal = "JHEP",
    volume = "11",
    pages = "037",
    year = "2025"
}

@article{Lee:2024wes,
    author = "Lee, Hyun Min and Park, Myeonghun and Sanz, Veronica",
    title = "{Gravity-Mediated Dark Matter at a low reheating temperature}",
    eprint = "2412.07850",
    archivePrefix = "arXiv",
    primaryClass = "hep-ph",
    doi = "10.1007/JHEP05(2025)126",
    journal = "JHEP",
    volume = "05",
    pages = "126",
    year = "2025"
}

@article{deGiorgi:2024elx,
    author = "de Giorgi, Arturo and Ramos, Maria",
    title = "{Extra-dimensional axion patterns}",
    eprint = "2412.00179",
    archivePrefix = "arXiv",
    primaryClass = "hep-ph",
    reportNumber = "IPPP/24/75, CERN-TH-2024-205",
    doi = "10.1103/PhysRevD.111.075006",
    journal = "Phys. Rev. D",
    volume = "111",
    number = "7",
    pages = "075006",
    year = "2025"
}

@article{Chivukula:2024nzt,
    author = "Chivukula, R. Sekhar and Gill, Joshua A. and Mohan, Kirtimaan A. and Sanamyan, George and Sengupta, Dipan and Simmons, Elizabeth H. and Wang, Xing",
    title = "{Limits on Kaluza-Klein portal dark matter models}",
    eprint = "2411.02509",
    archivePrefix = "arXiv",
    primaryClass = "hep-ph",
    doi = "10.1103/PhysRevD.111.075030",
    journal = "Phys. Rev. D",
    volume = "111",
    number = "7",
    pages = "075030",
    year = "2025"
}

@article{Koutroulis:2024wjl,
    author = "Koutroulis, Fotis and Megias, Eugenio and Pokorski, Stefan and Quiros, Mariano",
    title = "{Dark branes for dark matter}",
    eprint = "2403.06276",
    archivePrefix = "arXiv",
    primaryClass = "hep-ph",
    doi = "10.1103/PhysRevD.110.055015",
    journal = "Phys. Rev. D",
    volume = "110",
    number = "5",
    pages = "055015",
    year = "2024"
}

@article{Voronchikhin:2023qig,
    author = "Voronchikhin, I. V. and Kirpichnikov, D. V.",
    title = "{Probing scalar, Dirac, Majorana, and vector dark matter through a spin-0 electron-specific mediator at electron fixed-target experiments}",
    eprint = "2312.15697",
    archivePrefix = "arXiv",
    primaryClass = "hep-ph",
    doi = "10.1103/PhysRevD.109.075012",
    journal = "Phys. Rev. D",
    volume = "109",
    number = "7",
    pages = "075012",
    year = "2024"
}

@article{Chen:2023oip,
    author = "Chen, Qing and Zhou, Shuang-Yong",
    title = "{Bigravity portal dark matter}",
    eprint = "2310.03079",
    archivePrefix = "arXiv",
    primaryClass = "hep-ph",
    reportNumber = "USTC-ICTS/PCFT-23-30",
    doi = "10.1103/PhysRevD.109.095035",
    journal = "Phys. Rev. D",
    volume = "109",
    number = "9",
    pages = "095035",
    year = "2024"
}

@article{Voronchikhin:2023znz,
    author = "Voronchikhin, I. V. and Kirpichnikov, D. V.",
    title = "{Resonant probing spin-0 and spin-2 dark matter mediators with fixed target experiments}",
    eprint = "2304.14052",
    archivePrefix = "arXiv",
    primaryClass = "hep-ph",
    doi = "10.1103/PhysRevD.107.115034",
    journal = "Phys. Rev. D",
    volume = "107",
    number = "11",
    pages = "115034",
    year = "2023"
}

@article{Bhattacharya:2024wpy,
    author = "Bhattacharya, Srimoy and Choudhury, Debajyoti and Maharana, Suvam and Srivastava, Tripurari",
    title = "{Broad axion resonances: Clockwork ALPs at hadron colliders}",
    eprint = "2409.05983",
    archivePrefix = "arXiv",
    primaryClass = "hep-ph",
    doi = "10.1103/dsmk-v7n8",
    journal = "Phys. Rev. D",
    volume = "112",
    number = "5",
    pages = "055030",
    year = "2025"
}

@article{Choi:2014rja,
    author = "Choi, Kiwoon and Kim, Hyungjin and Yun, Seokhoon",
    title = "{Natural inflation with multiple sub-Planckian axions}",
    eprint = "1404.6209",
    archivePrefix = "arXiv",
    primaryClass = "hep-th",
    reportNumber = "CTPU-14-04",
    doi = "10.1103/PhysRevD.90.023545",
    journal = "Phys. Rev. D",
    volume = "90",
    pages = "023545",
    year = "2014"
}

@article{Choi:2015fiu,
    author = "Choi, Kiwoon and Im, Sang Hui",
    title = "{Realizing the relaxion from multiple axions and its UV completion with high scale supersymmetry}",
    eprint = "1511.00132",
    archivePrefix = "arXiv",
    primaryClass = "hep-ph",
    reportNumber = "CTPU-15-16",
    doi = "10.1007/JHEP01(2016)149",
    journal = "JHEP",
    volume = "01",
    pages = "149",
    year = "2016"
}

@article{Kaplan:2015fuy,
    author = "Kaplan, David E. and Rattazzi, Riccardo",
    title = "{Large field excursions and approximate discrete symmetries from a clockwork axion}",
    eprint = "1511.01827",
    archivePrefix = "arXiv",
    primaryClass = "hep-ph",
    doi = "10.1103/PhysRevD.93.085007",
    journal = "Phys. Rev. D",
    volume = "93",
    number = "8",
    pages = "085007",
    year = "2016"
}

@article{Will:2014kxa,
    author = "Will, Clifford M.",
    title = "{The Confrontation between General Relativity and Experiment}",
    eprint = "1403.7377",
    archivePrefix = "arXiv",
    primaryClass = "gr-qc",
    doi = "10.12942/lrr-2014-4",
    journal = "Living Rev. Rel.",
    volume = "17",
    pages = "4",
    year = "2014"
}

@article{Fierz:1939ix,
    author = "Fierz, M. and Pauli, W.",
    title = "{On relativistic wave equations for particles of arbitrary spin in an electromagnetic field}",
    doi = "10.1098/rspa.1939.0140",
    journal = "Proc. Roy. Soc. Lond. A",
    volume = "173",
    pages = "211--232",
    year = "1939"
}

@article{LIGOScientific:2020tif,
    author = "Abbott, R. and others",
    collaboration = "LIGO Scientific, Virgo",
    title = "{Tests of general relativity with binary black holes from the second LIGO-Virgo gravitational-wave transient catalog}",
    eprint = "2010.14529",
    archivePrefix = "arXiv",
    primaryClass = "gr-qc",
    reportNumber = "LIGO-P2000091",
    doi = "10.1103/PhysRevD.103.122002",
    journal = "Phys. Rev. D",
    volume = "103",
    number = "12",
    pages = "122002",
    year = "2021"
}

@article{Bernus:2020szc,
    author = "Bernus, L. and Minazzoli, O. and Fienga, A. and Gastineau, M. and Laskar, J. and Deram, P. and Di Ruscio, A.",
    title = "{Constraint on the Yukawa suppression of the Newtonian potential from the planetary ephemeris INPOP19a}",
    eprint = "2006.12304",
    archivePrefix = "arXiv",
    primaryClass = "gr-qc",
    doi = "10.1103/PhysRevD.102.021501",
    journal = "Phys. Rev. D",
    volume = "102",
    number = "2",
    pages = "021501",
    year = "2020"
}

@article{Bellazzini:2023nqj,
    author = "Bellazzini, Brando and Isabella, Giulia and Ricossa, Sergio and Riva, Francesco",
    title = "{Massive gravity is not positive}",
    eprint = "2304.02550",
    archivePrefix = "arXiv",
    primaryClass = "hep-th",
    doi = "10.1103/PhysRevD.109.024051",
    journal = "Phys. Rev. D",
    volume = "109",
    number = "2",
    pages = "024051",
    year = "2024"
}

@article{Dong:2025dpy,
    author = "Dong, Ziyu and Jeong, Jaehoon and Pomarol, Alex",
    title = "{Causal bounds on EFTs with anomalies with a pseudoscalar, photons, and gravitons}",
    eprint = "2510.12138",
    archivePrefix = "arXiv",
    primaryClass = "hep-th",
    reportNumber = "KIAS-Q25016",
    doi = "10.1007/JHEP02(2026)102",
    journal = "JHEP",
    volume = "02",
    pages = "102",
    year = "2026"
}

@article{deRham:2010ik,
    author = "de Rham, Claudia and Gabadadze, Gregory",
    title = "{Generalization of the Fierz-Pauli Action}",
    eprint = "1007.0443",
    archivePrefix = "arXiv",
    primaryClass = "hep-th",
    reportNumber = "NYU-TH-06-13-10",
    doi = "10.1103/PhysRevD.82.044020",
    journal = "Phys. Rev. D",
    volume = "82",
    pages = "044020",
    year = "2010"
}

@article{deRham:2010kj,
    author = "de Rham, Claudia and Gabadadze, Gregory and Tolley, Andrew J.",
    title = "{Resummation of Massive Gravity}",
    eprint = "1011.1232",
    archivePrefix = "arXiv",
    primaryClass = "hep-th",
    doi = "10.1103/PhysRevLett.106.231101",
    journal = "Phys. Rev. Lett.",
    volume = "106",
    pages = "231101",
    year = "2011"
}

@article{Hassan:2011zd,
    author = "Hassan, S. F. and Rosen, Rachel A.",
    title = "{Bimetric Gravity from Ghost-free Massive Gravity}",
    eprint = "1109.3515",
    archivePrefix = "arXiv",
    primaryClass = "hep-th",
    doi = "10.1007/JHEP02(2012)126",
    journal = "JHEP",
    volume = "02",
    pages = "126",
    year = "2012"
}

@article{Hassan:2011vm,
    author = "Hassan, S. F. and Rosen, Rachel A.",
    title = "{On Non-Linear Actions for Massive Gravity}",
    eprint = "1103.6055",
    archivePrefix = "arXiv",
    primaryClass = "hep-th",
    doi = "10.1007/JHEP07(2011)009",
    journal = "JHEP",
    volume = "07",
    pages = "009",
    year = "2011"
}

@article{deRham:2013tfa,
    author = "de Rham, Claudia and Matas, Andrew and Tolley, Andrew J.",
    title = "{New Kinetic Interactions for Massive Gravity?}",
    eprint = "1311.6485",
    archivePrefix = "arXiv",
    primaryClass = "hep-th",
    doi = "10.1088/0264-9381/31/16/165004",
    journal = "Class. Quant. Grav.",
    volume = "31",
    pages = "165004",
    year = "2014"
}

@article{Arkani-Hamed:2002bjr,
    author = "Arkani-Hamed, Nima and Georgi, Howard and Schwartz, Matthew D.",
    title = "{Effective field theory for massive gravitons and gravity in theory space}",
    eprint = "hep-th/0210184",
    archivePrefix = "arXiv",
    reportNumber = "HUTP-02-A051",
    doi = "10.1016/S0003-4916(03)00068-X",
    journal = "Annals Phys.",
    volume = "305",
    pages = "96--118",
    year = "2003"
}

@article{Schwartz:2003vj,
    author = "Schwartz, Matthew D.",
    title = "{Constructing gravitational dimensions}",
    eprint = "hep-th/0303114",
    archivePrefix = "arXiv",
    reportNumber = "HUTP-03-A021",
    doi = "10.1103/PhysRevD.68.024029",
    journal = "Phys. Rev. D",
    volume = "68",
    pages = "024029",
    year = "2003"
}

@article{Falkowski:2020mjq,
    author = "Falkowski, Adam and Isabella, Giulia",
    title = "{Matter coupling in massive gravity}",
    eprint = "2001.06800",
    archivePrefix = "arXiv",
    primaryClass = "hep-th",
    doi = "10.1007/JHEP04(2020)014",
    journal = "JHEP",
    volume = "04",
    pages = "014",
    year = "2020"
}

@article{Bonifacio:2019mgk,
    author = "Bonifacio, James and Hinterbichler, Kurt and Rosen, Rachel A.",
    title = "{Constraints on a gravitational Higgs mechanism}",
    eprint = "1903.09643",
    archivePrefix = "arXiv",
    primaryClass = "hep-th",
    doi = "10.1103/PhysRevD.100.084017",
    journal = "Phys. Rev. D",
    volume = "100",
    number = "8",
    pages = "084017",
    year = "2019"
}

@article{Bonifacio:2019ioc,
    author = "Bonifacio, James and Hinterbichler, Kurt",
    title = "{Unitarization from Geometry}",
    eprint = "1910.04767",
    archivePrefix = "arXiv",
    primaryClass = "hep-th",
    doi = "10.1007/JHEP12(2019)165",
    journal = "JHEP",
    volume = "12",
    pages = "165",
    year = "2019"
}

@article{Chivukula:2020hvi,
    author = "Chivukula, R. Sekhar and Foren, Dennis and Mohan, Kirtimaan A. and Sengupta, Dipan and Simmons, Elizabeth H.",
    title = "{Massive Spin-2 Scattering Amplitudes in Extra-Dimensional Theories}",
    eprint = "2002.12458",
    archivePrefix = "arXiv",
    primaryClass = "hep-ph",
    reportNumber = "MSUHEP-20-003",
    doi = "10.1103/PhysRevD.101.075013",
    journal = "Phys. Rev. D",
    volume = "101",
    number = "7",
    pages = "075013",
    year = "2020"
}

@article{deGiorgi:2020qlg,
    author = "de Giorgi, A. and Vogl, S.",
    title = "{Unitarity in KK-graviton production: A case study in warped extra-dimensions}",
    eprint = "2012.09672",
    archivePrefix = "arXiv",
    primaryClass = "hep-ph",
    doi = "10.1007/JHEP04(2021)143",
    journal = "JHEP",
    volume = "04",
    pages = "143",
    year = "2021"
}

@article{Chivukula:2023qrt,
    author = "Chivukula, R. Sekhar and Gill, Joshua A. and Mohan, Kirtimaan A. and Sengupta, Dipan and Simmons, Elizabeth H. and Wang, Xing",
    title = "{Symmetries, spin-2 scattering amplitudes, and equivalence theorems in warped five-dimensional gravitational theories}",
    eprint = "2312.08576",
    archivePrefix = "arXiv",
    primaryClass = "hep-ph",
    reportNumber = "ADP23-28/T1237, MSUHEP-23-031",
    doi = "10.1103/PhysRevD.109.075016",
    journal = "Phys. Rev. D",
    volume = "109",
    number = "7",
    pages = "075016",
    year = "2024"
}

@article{Chivukula:2023sua,
    author = "Chivukula, R. Sekhar and Gill, Joshua A. and Mohan, Kirtimaan A. and Sengupta, Dipan and Simmons, Elizabeth H. and Wang, Xing",
    title = "{Scattering amplitudes of massive spin-2 Kaluza-Klein states with matter}",
    eprint = "2311.00770",
    archivePrefix = "arXiv",
    primaryClass = "hep-ph",
    reportNumber = "ADP-23-26/T1235, MSUHEP-23-029",
    doi = "10.1103/PhysRevD.109.015033",
    journal = "Phys. Rev. D",
    volume = "109",
    number = "1",
    pages = "015033",
    year = "2024"
}

@article{deGiorgi:2023mdy,
    author = "de Giorgi, A. and Vogl, S.",
    title = "{Gravity-matter sum rules in models with a single extra-dimension}",
    eprint = "2311.01507",
    archivePrefix = "arXiv",
    primaryClass = "hep-ph",
    reportNumber = "IFT-UAM/CSIC-23-139",
    doi = "10.1007/JHEP05(2024)315",
    journal = "JHEP",
    volume = "05",
    pages = "315",
    year = "2024"
}

@article{ATLAS:2012yve,
    author = "Aad, Georges and others",
    collaboration = "ATLAS",
    title = "{Observation of a new particle in the search for the Standard Model Higgs boson with the ATLAS detector at the LHC}",
    eprint = "1207.7214",
    archivePrefix = "arXiv",
    primaryClass = "hep-ex",
    reportNumber = "CERN-PH-EP-2012-218",
    doi = "10.1016/j.physletb.2012.08.020",
    journal = "Phys. Lett. B",
    volume = "716",
    pages = "1--29",
    year = "2012"
}

@article{CMS:2012qbp,
    author = "Chatrchyan, Serguei and others",
    collaboration = "CMS",
    title = "{Observation of a New Boson at a Mass of 125 GeV with the CMS Experiment at the LHC}",
    eprint = "1207.7235",
    archivePrefix = "arXiv",
    primaryClass = "hep-ex",
    reportNumber = "CMS-HIG-12-028, CERN-PH-EP-2012-220",
    doi = "10.1016/j.physletb.2012.08.021",
    journal = "Phys. Lett. B",
    volume = "716",
    pages = "30--61",
    year = "2012"
}

@article{Choi:2026kxu,
    author = "Choi, Kiwoon and Lee, Chang Hyeon and Shin, Chang Sub",
    title = "{Axion Quality in Warped Extra-Dimension}",
    eprint = "2604.08700",
    archivePrefix = "arXiv",
    primaryClass = "hep-ph",
    month = "4",
    year = "2026"
}

@article{Choi:2003wr,
    author = "Choi, Ki-woon",
    title = "{A QCD axion from higher dimensional gauge field}",
    eprint = "hep-ph/0308024",
    archivePrefix = "arXiv",
    reportNumber = "KAIST-TH-2003-07",
    doi = "10.1103/PhysRevLett.92.101602",
    journal = "Phys. Rev. Lett.",
    volume = "92",
    pages = "101602",
    year = "2004"
}

@article{Reece:2025thc,
    author = "Reece, Matthew",
    title = "{Extra-dimensional axion expectations}",
    eprint = "2406.08543",
    archivePrefix = "arXiv",
    primaryClass = "hep-ph",
    doi = "10.1007/JHEP07(2025)130",
    journal = "JHEP",
    volume = "07",
    pages = "130",
    year = "2025"
}

@article{Craig:2024dnl,
    author = "Craig, Nathaniel and Kongsore, Marius",
    title = "{High-quality axions from higher-form symmetries in extra dimensions}",
    eprint = "2408.10295",
    archivePrefix = "arXiv",
    primaryClass = "hep-ph",
    doi = "10.1103/PhysRevD.111.015047",
    journal = "Phys. Rev. D",
    volume = "111",
    number = "1",
    pages = "015047",
    year = "2025"
}

@article{Dienes:1998sb,
    author = "Dienes, Keith R. and Dudas, Emilian and Gherghetta, Tony",
    title = "{Neutrino oscillations without neutrino masses or heavy mass scales: A Higher dimensional seesaw mechanism}",
    eprint = "hep-ph/9811428",
    archivePrefix = "arXiv",
    reportNumber = "CERN-TH-98-370",
    doi = "10.1016/S0550-3213(99)00377-6",
    journal = "Nucl. Phys. B",
    volume = "557",
    pages = "25",
    year = "1999"
}

@article{Arkani-Hamed:1998wuz,
    author = "Arkani-Hamed, Nima and Dimopoulos, Savas and Dvali, G. R. and March-Russell, John",
    title = "{Neutrino masses from large extra dimensions}",
    eprint = "hep-ph/9811448",
    archivePrefix = "arXiv",
    reportNumber = "SLAC-PUB-8014, SU-ITP-98-64",
    doi = "10.1103/PhysRevD.65.024032",
    journal = "Phys. Rev. D",
    volume = "65",
    pages = "024032",
    year = "2001"
}

@article{Dvali:1999cn,
    author = "Dvali, G. R. and Smirnov, Alexei Yu.",
    title = "{Probing large extra dimensions with neutrinos}",
    eprint = "hep-ph/9904211",
    archivePrefix = "arXiv",
    reportNumber = "NYU-TH-99-3-03",
    doi = "10.1016/S0550-3213(99)00574-X",
    journal = "Nucl. Phys. B",
    volume = "563",
    pages = "63--81",
    year = "1999"
}

@article{Lukas:2000rg,
    author = "Lukas, Andre and Ramond, Pierre and Romanino, Andrea and Ross, Graham G.",
    title = "{Neutrino Masses and Mixing in Brane World Theories}",
    eprint = "hep-ph/0011295",
    archivePrefix = "arXiv",
    reportNumber = "FERMILAB-PUB-00-284-T, OUTP-00-39P, SUSX-TH-00-018, UFIFT-HET-00-27",
    doi = "10.1088/1126-6708/2001/04/010",
    journal = "JHEP",
    volume = "04",
    pages = "010",
    year = "2001"
}

@article{Antoniadis:1998ig,
    author = "Antoniadis, Ignatios and Arkani-Hamed, Nima and Dimopoulos, Savas and Dvali, G. R.",
    title = "{New dimensions at a millimeter to a Fermi and superstrings at a TeV}",
    eprint = "hep-ph/9804398",
    archivePrefix = "arXiv",
    reportNumber = "SLAC-PUB-7801, SU-ITP-98-28, CPTH-S608-0498, IC-98-39",
    doi = "10.1016/S0370-2693(98)00860-0",
    journal = "Phys. Lett. B",
    volume = "436",
    pages = "257--263",
    year = "1998"
}

@article{Grossman:1999ra,
    author = "Grossman, Yuval and Neubert, Matthias",
    title = "{Neutrino masses and mixings in nonfactorizable geometry}",
    eprint = "hep-ph/9912408",
    archivePrefix = "arXiv",
    reportNumber = "CLNS-99-1656, SLAC-PUB-8330",
    doi = "10.1016/S0370-2693(00)00054-X",
    journal = "Phys. Lett. B",
    volume = "474",
    pages = "361--371",
    year = "2000"
}

@article{Huber:2003sf,
    author = "Huber, Stephan J. and Shafi, Qaisar",
    title = "{Seesaw mechanism in warped geometry}",
    eprint = "hep-ph/0309252",
    archivePrefix = "arXiv",
    reportNumber = "DESY-03-110, BA-03-11",
    doi = "10.1016/j.physletb.2003.12.012",
    journal = "Phys. Lett. B",
    volume = "583",
    pages = "293--303",
    year = "2004"
}

@article{Fong:2011xh,
    author = "Fong, Chee Sheng and Mohapatra, Rabindra N. and Sung, Ilmo",
    title = "{Majorana Neutrinos from Inverse Seesaw in Warped Extra Dimension}",
    eprint = "1107.4086",
    archivePrefix = "arXiv",
    primaryClass = "hep-ph",
    reportNumber = "YITP-SB-11-26, UM-DOE-ER-40762-504",
    doi = "10.1016/j.physletb.2011.08.069",
    journal = "Phys. Lett. B",
    volume = "704",
    pages = "171--178",
    year = "2011"
}

@article{Dienes:1999gw,
    author = "Dienes, Keith R. and Dudas, Emilian and Gherghetta, Tony",
    title = "{Invisible axions and large radius compactifications}",
    eprint = "hep-ph/9912455",
    archivePrefix = "arXiv",
    reportNumber = "CERN-TH-99-333",
    doi = "10.1103/PhysRevD.62.105023",
    journal = "Phys. Rev. D",
    volume = "62",
    pages = "105023",
    year = "2000"
}

@article{DiLella:2000dn,
    author = "Di Lella, L. and Pilaftsis, A. and Raffelt, G. and Zioutas, K.",
    title = "{Search for solar Kaluza-Klein axions in theories of low scale quantum gravity}",
    eprint = "hep-ph/0006327",
    archivePrefix = "arXiv",
    reportNumber = "WUE-ITP-2000-015",
    doi = "10.1103/PhysRevD.62.125011",
    journal = "Phys. Rev. D",
    volume = "62",
    pages = "125011",
    year = "2000"
}

@article{Flacke:2006ad,
    author = "Flacke, Thomas and Gripaios, Ben and March-Russell, John and Maybury, David",
    title = "{Warped axions}",
    eprint = "hep-ph/0611278",
    archivePrefix = "arXiv",
    reportNumber = "OUTP-06-16-P",
    doi = "10.1088/1126-6708/2007/01/061",
    journal = "JHEP",
    volume = "01",
    pages = "061",
    year = "2007"
}

@article{Albertus:2026fbe,
    author = "Albertus, Conrado and others",
    title = "{WISPedia -- the WISPs Encyclopedia}",
    eprint = "2602.09089",
    archivePrefix = "arXiv",
    primaryClass = "hep-ph",
    reportNumber = "IPPP/26/14, IFT-UAM/CSIC-26-9",
    month = "2",
    year = "2026"
}

@article{Arza:2026rsl,
    author = "Arza, A. and others",
    title = "{The COSMIC WISPers White Paper: The physics case for Weakly Interacting Slim Particles}",
    eprint = "2603.03433",
    archivePrefix = "arXiv",
    primaryClass = "hep-ph",
    reportNumber = "BARI-TH/784-26, CERN-TH-2026-016, IPPP/26/13, IFT-UAM/CSIC-26-13, KCL-PH-TH/2026-04, KEK-Cosmo-0411, KEK-TH-2804, LAPTH-008/26, MPP-2026-21, RESCEU-5/26, SLAC-PUB-260219, ST/T006994/1, ST/Y004531/1",
    month = "3",
    year = "2026"
}

@article{ATLAS:2012hvw,
    author = "Aad, Georges and others",
    collaboration = "ATLAS",
    title = "{Search for Extra Dimensions in diphoton events using proton-proton collisions recorded at $\sqrt{s}=7$ TeV with the ATLAS detector at the LHC}",
    eprint = "1210.8389",
    archivePrefix = "arXiv",
    primaryClass = "hep-ex",
    reportNumber = "CERN-PH-EP-2012-289",
    doi = "10.1088/1367-2630/15/4/043007",
    journal = "New J. Phys.",
    volume = "15",
    pages = "043007",
    year = "2013"
}

@article{ATLAS:2015shg,
    author = "Aad, Georges and others",
    collaboration = "ATLAS",
    title = "{Search for high-mass diphoton resonances in $pp$ collisions at $\sqrt{s}=8$ TeV with the ATLAS detector}",
    eprint = "1504.05511",
    archivePrefix = "arXiv",
    primaryClass = "hep-ex",
    reportNumber = "CERN-PH-EP-2015-043",
    doi = "10.1103/PhysRevD.92.032004",
    journal = "Phys. Rev. D",
    volume = "92",
    number = "3",
    pages = "032004",
    year = "2015"
}

@article{CMS:2016kgr,
    author = "Khachatryan, Vardan and others",
    collaboration = "CMS",
    title = "{Search for high-mass diphoton resonances in proton{\textendash}proton collisions at 13 TeV and combination with 8 TeV search}",
    eprint = "1609.02507",
    archivePrefix = "arXiv",
    primaryClass = "hep-ex",
    reportNumber = "CMS-EXO-16-027, CERN-EP-2016-216, CMS-EXO-16-02",
    doi = "10.1016/j.physletb.2017.01.027",
    journal = "Phys. Lett. B",
    volume = "767",
    pages = "147--170",
    year = "2017"
}

@article{CDF:2010muc,
    author = "Aaltonen, T. and others",
    collaboration = "CDF",
    title = "{Search for Randall-Sundrum Gravitons in the Diphoton Channel at CDF}",
    eprint = "1012.2795",
    archivePrefix = "arXiv",
    primaryClass = "hep-ex",
    reportNumber = "FERMILAB-PUB-10-498-E",
    doi = "10.1103/PhysRevD.83.011102",
    journal = "Phys. Rev. D",
    volume = "83",
    pages = "011102",
    year = "2011"
}

@article{D0:2005srl,
    author = "Abazov, V. M. and others",
    collaboration = "D0",
    title = "{Search for Randall-Sundrum gravitons in dilepton and diphoton final states}",
    eprint = "hep-ex/0505018",
    archivePrefix = "arXiv",
    reportNumber = "FERMILAB-PUB-05-126-E-T",
    doi = "10.1103/PhysRevLett.95.091801",
    journal = "Phys. Rev. Lett.",
    volume = "95",
    pages = "091801",
    year = "2005"
}

@article{ATLAS:2011ab,
    author = "Aad, Georges and others",
    collaboration = "ATLAS",
    title = "{Search for extra dimensions using diphoton events in 7 TeV proton{\textendash}proton collisions with the ATLAS detector}",
    eprint = "1112.2194",
    archivePrefix = "arXiv",
    primaryClass = "hep-ex",
    reportNumber = "CERN-PH-EP-2011-189",
    doi = "10.1016/j.physletb.2012.03.022",
    journal = "Phys. Lett. B",
    volume = "710",
    pages = "538--556",
    year = "2012"
}

@article{ATLAS:2011tzr,
    author = "Aad, Georges and others",
    collaboration = "ATLAS",
    title = "{Search for dilepton resonances in $pp$ collisions at $\sqrt{s}=7$ TeV with the ATLAS detector}",
    eprint = "1108.1582",
    archivePrefix = "arXiv",
    primaryClass = "hep-ex",
    reportNumber = "CERN-PH-EP-2011-123",
    doi = "10.1103/PhysRevLett.107.272002",
    journal = "Phys. Rev. Lett.",
    volume = "107",
    pages = "272002",
    year = "2011"
}

@article{CMS:2011bsw,
    author = "Chatrchyan, Serguei and others",
    collaboration = "CMS",
    title = "{Search for signatures of extra dimensions in the diphoton mass spectrum at the Large Hadron Collider}",
    eprint = "1112.0688",
    archivePrefix = "arXiv",
    primaryClass = "hep-ex",
    reportNumber = "CERN-PH-EP-2011-173, CMS-EXO-11-038",
    doi = "10.1103/PhysRevLett.108.111801",
    journal = "Phys. Rev. Lett.",
    volume = "108",
    pages = "111801",
    year = "2012"
}

@article{CMS:2019qem,
    author = "Sirunyan, Albert M and others",
    collaboration = "CMS",
    title = "{A multi-dimensional search for new heavy resonances decaying to boosted WW, WZ, or ZZ boson pairs in the dijet final state at 13 TeV}",
    eprint = "1906.05977",
    archivePrefix = "arXiv",
    primaryClass = "hep-ex",
    reportNumber = "CMS-B2G-18-002, CERN-EP-2019-107",
    doi = "10.1140/epjc/s10052-020-7773-5",
    journal = "Eur. Phys. J. C",
    volume = "80",
    number = "3",
    pages = "237",
    year = "2020"
}

@article{CMS:2021roc,
    author = "Tumasyan, Armen and others",
    collaboration = "CMS",
    title = "{Search for heavy resonances decaying to a pair of Lorentz-boosted Higgs bosons in final states with leptons and a bottom quark pair at $ \sqrt{s} $= 13 TeV}",
    eprint = "2112.03161",
    archivePrefix = "arXiv",
    primaryClass = "hep-ex",
    reportNumber = "CMS-B2G-20-007, CERN-EP-2021-226",
    doi = "10.1007/JHEP05(2022)005",
    journal = "JHEP",
    volume = "05",
    pages = "005",
    year = "2022"
}

@article{CMS:2019kaf,
    author = "Sirunyan, Albert M and others",
    collaboration = "CMS",
    title = "{Combination of CMS searches for heavy resonances decaying to pairs of bosons or leptons}",
    eprint = "1906.00057",
    archivePrefix = "arXiv",
    primaryClass = "hep-ex",
    reportNumber = "CMS-B2G-18-006, CERN-EP-2019-110",
    doi = "10.1016/j.physletb.2019.134952",
    journal = "Phys. Lett. B",
    volume = "798",
    pages = "134952",
    year = "2019"
}

@article{ATLAS:2020tlo,
    author = "Aad, Georges and others",
    collaboration = "ATLAS",
    title = "{Search for heavy resonances decaying into a pair of Z bosons in the $\ell ^+\ell ^-\ell '^+\ell '^-$ and $\ell ^+\ell ^-\nu {{\bar{\nu }}}$ final states using 139 $\mathrm {fb}^{-1}$ of proton{\textendash}proton collisions at $\sqrt{s} = 13\,$TeV with the ATLAS detector}",
    eprint = "2009.14791",
    archivePrefix = "arXiv",
    primaryClass = "hep-ex",
    reportNumber = "CERN-EP-2020-153",
    doi = "10.1140/epjc/s10052-021-09013-y",
    journal = "Eur. Phys. J. C",
    volume = "81",
    number = "4",
    pages = "332",
    year = "2021"
}

@article{ATLAS:2017uhp,
    author = "Aaboud, Morad and others",
    collaboration = "ATLAS",
    title = "{Search for heavy resonances decaying into $WW$ in the $e\nu\mu\nu$ final state in $pp$ collisions at $\sqrt{s}=13$ TeV with the ATLAS detector}",
    eprint = "1710.01123",
    archivePrefix = "arXiv",
    primaryClass = "hep-ex",
    reportNumber = "CERN-EP-2017-214",
    doi = "10.1140/epjc/s10052-017-5491-4",
    journal = "Eur. Phys. J. C",
    volume = "78",
    number = "1",
    pages = "24",
    year = "2018"
}

@article{Bierlich:2022pfr,
    author = "Bierlich, Christian and others",
    title = "{A comprehensive guide to the physics and usage of PYTHIA 8.3}",
    eprint = "2203.11601",
    archivePrefix = "arXiv",
    primaryClass = "hep-ph",
    reportNumber = "LU-TP 22-16, MCNET-22-04, FERMILAB-PUB-22-227-SCD",
    doi = "10.21468/SciPostPhysCodeb.8",
    journal = "SciPost Phys. Codeb.",
    volume = "2022",
    pages = "8",
    year = "2022"
}

@article{Bijnens:2001gh,
    author = "Bijnens, J. and Eerola, P. and Maul, M. and Mansson, A. and Sjostrand, T.",
    title = "{QCD signatures of narrow graviton resonances in hadron colliders}",
    eprint = "hep-ph/0101316",
    archivePrefix = "arXiv",
    reportNumber = "LU-TP-01-05",
    doi = "10.1016/S0370-2693(01)00238-6",
    journal = "Phys. Lett. B",
    volume = "503",
    pages = "341--348",
    year = "2001"
}

@article{dEnterria:2023npy,
    author = "d'Enterria, David and Tamlihat, Malak Ait and Schoeffel, Laurent and Shao, Hua-Sheng and Tayalati, Yahya",
    title = "{Collider constraints on massive gravitons coupling to photons}",
    eprint = "2306.15558",
    archivePrefix = "arXiv",
    primaryClass = "hep-ph",
    doi = "10.1016/j.physletb.2023.138237",
    journal = "Phys. Lett. B",
    volume = "846",
    pages = "138237",
    year = "2023"
}

@article{Garcia-Cely:2025ula,
    author = "Garc{\'\i}a-Cely, Camilo and Ringwald, Andreas",
    title = "{Stellar bounds on light spin-2 particles in bimetric theories}",
    eprint = "2511.03707",
    archivePrefix = "arXiv",
    primaryClass = "hep-ph",
    doi = "10.1088/1475-7516/2026/03/049",
    journal = "JCAP",
    volume = "03",
    pages = "049",
    year = "2026"
}

@article{Cembranos:2026cqk,
    author = "Cembranos, Jose A. R. and Cendal, {\'A}lvaro and Villarrubia-Rojo, Hector",
    title = "{Effects of massive spin-2 fields on gravitational wave propagation}",
    eprint = "2601.15201",
    archivePrefix = "arXiv",
    primaryClass = "gr-qc",
    reportNumber = "IPARCOS-UCM-26-002",
    doi = "10.1088/1475-7516/2026/06/008",
    journal = "JCAP",
    volume = "06",
    pages = "008",
    year = "2026"
}

@article{Cembranos:2017vgi,
    author = "Cembranos, J. A. R. and Maroto, A. L. and Villarrubia-Rojo, H.",
    title = "{Constraints on hidden gravitons from fifth-force experiments and stellar energy loss}",
    eprint = "1706.07818",
    archivePrefix = "arXiv",
    primaryClass = "hep-ph",
    doi = "10.1007/JHEP09(2017)104",
    journal = "JHEP",
    volume = "09",
    pages = "104",
    year = "2017"
}

@article{Cembranos:2021vdv,
    author = "Cembranos, J. A. R. and Delgado, R. L. and Villarrubia-Rojo, H.",
    title = "{LHC constraints on hidden gravitons}",
    eprint = "2108.00930",
    archivePrefix = "arXiv",
    primaryClass = "hep-ph",
    reportNumber = "TUM-EFT 146/21",
    doi = "10.1007/JHEP01(2022)129",
    journal = "JHEP",
    volume = "01",
    pages = "129",
    year = "2022"
}

@article{Gue:2026wqd,
    author = "Gu{\'e}, Jordan and d'Enterria, David",
    title = "{Bounds on massive graviton-like particles from searches for axion-like particles coupling to photons}",
    eprint = "2605.00549",
    archivePrefix = "arXiv",
    primaryClass = "hep-ph",
    month = "5",
    year = "2026"
}

@article{Armaleo:2019gil,
    author = "Armaleo, Juan Manuel and L{\'o}pez Nacir, Diana and Urban, Federico R.",
    title = "{Binary pulsars as probes for spin-2 ultralight dark matter}",
    eprint = "1909.13814",
    archivePrefix = "arXiv",
    primaryClass = "astro-ph.HE",
    doi = "10.1088/1475-7516/2020/01/053",
    journal = "JCAP",
    volume = "01",
    pages = "053",
    year = "2020"
}

@article{Armaleo:2020yml,
    author = "Armaleo, Juan Manuel and L{\'o}pez Nacir, Diana and Urban, Federico R.",
    title = "{Pulsar timing array constraints on spin-2 ULDM}",
    eprint = "2005.03731",
    archivePrefix = "arXiv",
    primaryClass = "astro-ph.CO",
    doi = "10.1088/1475-7516/2020/09/031",
    journal = "JCAP",
    volume = "09",
    pages = "031",
    year = "2020"
}

@article{Clarke:2026cdl,
    author = "Clarke, A. and Flambaum, V. V. and Pospelov, M. and Samsonov, I. B.",
    title = "{Signatures of gravity-mediated dark matter interaction in theories with large extra dimensions}",
    eprint = "2606.23178",
    archivePrefix = "arXiv",
    primaryClass = "hep-ph",
    month = "6",
    year = "2026"
}

@article{Voronchikhin:2025eqm,
    author = "Voronchikhin, I. V. and Kirpichnikov, D. V.",
    title = "{Examining scalar portal inelastic dark matter with lepton fixed-target experiments}",
    eprint = "2505.04290",
    archivePrefix = "arXiv",
    primaryClass = "hep-ph",
    doi = "10.1103/qvbp-rhsr",
    journal = "Phys. Rev. D",
    volume = "113",
    number = "1",
    pages = "015031",
    year = "2026"
}

@article{Voronchikhin:2024ygo,
    author = "Voronchikhin, I. V. and Kirpichnikov, D. V.",
    title = "{The bremsstrahlung-like production of the massive spin-2 dark matter mediator}",
    eprint = "2412.10150",
    archivePrefix = "arXiv",
    primaryClass = "hep-ph",
    doi = "10.1140/epjc/s10052-025-14868-6",
    journal = "Eur. Phys. J. C",
    volume = "85",
    number = "10",
    pages = "1110",
    year = "2025"
}

@misc{GravBounds,
author={de Giorgi, Arturo and Marcoli, Matteo and Silvetti, Federico },
title={GravBounds, v1.0},
howpublished={\url{https://gitlab.com/federico_silvetti/gravbounds/-/tree/release-1.0}},
year={2026}
}

@article{ATLAS:2019qmc,
    author = "Aad, Georges and others",
    collaboration = "ATLAS",
    title = "{Electron and photon performance measurements with the ATLAS detector using the 2015{\textendash}2017 LHC proton-proton collision data}",
    eprint = "1908.00005",
    archivePrefix = "arXiv",
    primaryClass = "hep-ex",
    reportNumber = "CERN-EP-2019-145",
    doi = "10.1088/1748-0221/14/12/P12006",
    journal = "JINST",
    volume = "14",
    number = "12",
    pages = "P12006",
    year = "2019"
}
\end{document}